\documentclass[a4paper,11pt]{article}
\pdfoutput=1 

\usepackage{jheppub} 

\usepackage[T1]{fontenc} 
\usepackage{hyperref}
\usepackage{ulem}

\title{\boldmath Light sterile neutrino and leptogenesis}

\author[a]{Ki-Young Jung,}
\author[a,1]{Kim Siyeon\note{Corresponding author.}}

\affiliation[a]{Department of Physics, Chung-Ang University, Seoul 06974 Korea}

\emailAdd{dmsrl1100@cau.ac.kr}
\emailAdd{siyeon@cau.ac.kr}

\abstract{We studied models of leptogenesis where three right-handed Majorana neutrinos are involved and the minimal-extended seesaw mechanism including an additional singlet field produces four light neutrinos. This study shows that the type of mass ordering and heavy Majorana scales can be determined by inputting the simplest orthogonal matrix into the Casas-Ibarra(CI) representation of seesaw. The CP asymmetry produced from the decays of heavy neutrinos and the dilution mass are predicted in terms of the mass and mixing elements of the fourth neutrino. Upon the choice of CI matrix, the existence of a light sterile neutrino is required to explain the high-energy lepton asymmetry in light of  phenomenological measurements. Although there are several free parameters attributable to an additional neutrino, the model can be in part constrained by low-energy experiments such as sterile neutrino searches and neutrinoless double-beta decays, as well as the observed baryon asymmetry in the universe. }

\arxivnumber{2205.13860}

\begin{document} 
\maketitle
\flushbottom

\section{Introduction}
\label{sec:intro}

Neutrino Physics is about to turn the corner on its way. Next-generation neutrino oscillation experiments under construction, such as DUNE\cite{DUNE:2020lwj} and HK\cite{Hyper-Kamiokande:2018ofw}, will measure the ordering of masses and CP violation phase in Potecorvo-Maki-Nakagawa-Sakata(PMNS) matrix \cite{Pontecorvo:1967fh}\cite{Maki:1962mu} to complete a puzzle board of three massive neutrinos. In addition to mass, mixing angles, and CP violations in the three-neutrino frame, there are other fundamental problems and efforts have been made to address those problems \cite{Agostini:2017jim}. 
For example, to understand whether a neutrino is a Majorana particle, a number of neutrinoless double beta decay experiments \cite{CUORE:2017tlq, GERDA:2018pmc, Majorana:2017csj, EXO-200:2017hwz, CUPID-0:2018rcs, Fischer:2018squ, NEXT:2015wlq, Alenkov:2019jis, Barabash:2017sxf} are being conducted, and next-generation experiments such as Legend\cite{LEGEND:2017cdu}, KamLAND2-Zen\cite{KamLAND-Zen:2016pfg}, and AMoRE-II\cite{AMoRE:2015asn} that can determine the effective mass of electron neutrino $m_{ee}$ with aggressive sensitivity will also be launched. 
Another issue is the mass-induced oscillations between active neutrinos and sterile neutrino. Short-baseline oscillation experiments using electron antineutrinos from reactors\cite{NEOS:2016wee, PROSPECT:2018dtt, DANSS:2018fnn, Serebrov:2020kmd, SoLid:2020cen, STEREO:2018rfh, RENO:2020hva} and those using muon neutrinos(antineutrinos) from accelerator-based beam lines\cite{LSND:2001aii, MiniBooNE:2018esg} were performed to find the mixing angles and mass of the fourth neutrino. Assuming that the mass of the fourth neutrino is on the order of 1 eV, more precise experiments such as SBN\cite{Machado:2019oxb} and $\mathrm{JSNS}^2$\cite{Ajimura:2020qni} are underway to obtain data in the near future.

Lepton number violation is generated from processes including Majorana neutrino. Majorana neutrinos that existed with masses near GUT scale in early universe could produce lepton asymmetry before sphaleron process at electroweak scale \cite{Kuzmin:1985mm}. Thus the baryongenesis via leptogenesis\cite{Fukugita:1986hr}\cite{Luty:1992un} is one of the natural explanation for the observed baryon asymmetry\cite{Planck:2018nkj}. 
As in Sakharov's three conditions for matter-antimatter asymmetry \cite{Sakharov:1967dj}, the generation of lepton asymmetry also requires sufficient CP violation in lepton decay, and a careful balance of the produced asymmetry and washout effect, which is done by comparing the decay width to the Hubble expansion rate \cite{Harvey:1990qw}. The CP-violating decay of a Majorana neutrino via Yukawa coupling requires the vertex contribution of one-loop level including different neutrinos\cite{Covi:1996wh, Roulet:1997xa, Nielsen:2001fy}. At least two right-hand(RH) neutrinos are needed for Yukawa matrix to contain a phase, and so the leptogenesis with two heavy neutrinos is called the minimal model \cite{Xing:2020ald, Kang:2021stv}. It is minimal also in a sense that the lepton asymmetry is described mostly in terms of measurable quantities at low energy. 
Previous works on minimal models \cite{Chang:2004wy, Siyeon:2016wro} motivated a model extension to three heavy neutrinos without approximation. If all scales of RH neutrinos of seesaw mechanism are higher than $10^{12}$ GeV, lepton flavor contribution in the decays of heavy neutrinos is excluded out of consideration for the leptogenesis \cite{Abada:2006ea, Nardi:2006fx, Moffat:2018smo}. The scales of three heavy neutrinos are assumed to be hierarchically different so that the resonance from the self-energy contribution is avoided \cite{Pilaftsis:2003gt}. 

Seesaw mechanism that connects low-energy sector to GUT-scale enables to furnish Yukawa matrix in terms of light masses, low-energy transformation elements, and furthermore heavy Majorana neutrinos. 
Canonical seesaw mechanism gives rise to three light masses by leveraging three active neutrinos up with two or three heavy ones. 
According to Casas-Ibarra(CI) expression, the complexity of a 3-by-3 Yukawa matrix cannot be generated unless the CI orthogonal matrix contains complex angles \cite{Casas:2001sr}. 
Models in this article include only real CI rotations for seesaw mechanism, and construct so-called extended Dirac matrix, a 4-by-3 matrix consisting of a 3-by-3 Dirac matrix and 1-by-3 Majorana matrix. 
In minimal extended seesaw(MES) model, additional Majorana mass terms are originated by the coupling of heavy Majorana neutrino with a lighter Majorana neutrino. If the additional masses are comparable to Dirac masses, they can be a part of the extended Dirac matrix and get involved in seesaw mechanism \cite{Rodejohann:2009ve, Barry:2011wb, Zhang:2011vh, Nath:2016mts,Goswami:2021eqy}. 
In this study, the complex Yukawa couplings are traced from a non-unitary partial block of the unitary 4-neutrino mixing matrix, unlike other approaches where the necessary complexity were traced from complex CI angles.

Currently available phenomenological constraints are (i) observed matter-antimatter asymmetry $Y_B$\cite{Planck:2018nkj}, (ii) effective electron-neutrino masses $m_{ee}$ in neutrinoless double-beta decays\cite{Giunti:2015kza}, and (iii) the elements of PMNS matrix and extended mixing elements with fourth neutrino from oscillation experiments\cite{Parke:2015goa}\cite{Gariazzo:2017fdh}. Both normal ordering(NO) and inverted ordering(IO) are yet candidates of mass ordering, while the Dirac phase in PMNS excluded zero from its range at $3\sigma$ CL \cite{T2K:2019bcf}\cite{NOvA:2019cyt}. 
Including one sterile neutrino to the light Majorana neutrino content doubles the number of mixing angles and the number of phases, respectively. It is difficult to test the accessibility of phenomenological observations because phases are added that are either too many or too complicate to measure with known methods. In the near future, we should explore the possibility to ruling out, at least in part, the domains of parameters through experiments that may yield new results one by one. 

The outline of contents is as follows:
Section II describes the current status of the phenomenology of Majorana neutrinos. 
Section III introduces a seesaw mechanism model for three neutrinos and MES for four neutrinos. The choice of CI orthogonal matrix determines whether the model is NO or IO.
Section IV validates that the model-generated lepton asymmetry is consistent with the observation of baryon asymmetry, and identifies the correlations between different mass scales.
In Section V, different phenomena $m_{ee}$ and $Y_B$ are estimated simultaneously in terms of combinations of phases, both Dirac and Majorana. Contours of $m_{ee}$ and $Y_B$ are drawn for the possibility that the ranges of those phases can be excluded in future experiments. It is followed by concluding remarks.

\section{Low-energy constraints: CP violation and Lepton number violation}
Three neutrinos that couple with weak bosons in the Standard Model(SM) are mixed states of the neutrinos defined by their masses. The unitary PMNS mixing matrix $V$ for 3 generations of neutrinos
is given by
    \begin{eqnarray}
    V &=& R\left(\theta_{23}\right)
          R\left(\theta_{13},\delta\right)
          R\left(\theta_{12}\right)
    \label{fulltrans}
    \end{eqnarray}
where each $R$ is a rotation matrix with a mixing angle $\theta_{ij}$ between $i$-th and $j$-th generations. According to the standard parametrization, the Dirac phase $\delta$ is combined with the smallest angle $\theta_{13}$ as in $\sin\theta_{13}e^{-i\delta}$ in the PMNS matrix. The detailed matrix form in terms of three angles and a phase is given in Eq.(\ref{ckm}).
Under assumption that neutrinos are Majorana particles, a diagonal phase transformation $P_2$ in Eq.(\ref{2phases}) is attached to $V$ such that
the Majorana phases $\eta_1$ and $\eta_2$ can be a part of the mass matrix of light neutrinos in the following way:
    \begin{equation}
    M_{\nu} = V^* \mathrm{Diag}(\check{m}_1,\check{m}_2, m_3) V^\dagger,
    \label{umu}
    \end{equation}
where $\check{m}_1 \equiv m_1e^{-2i\eta_1}$ and $\check{m}_2 \equiv m_2e^{-2i\eta_2}$. 
The notations and conventions for masses and transformation matrices in this work are taken from PDG \cite{ParticleDataGroup:2020ssz}. The PMNS matrix is then expressed as $U\equiv VP_2$.

Neutrinoless double-beta decay $0\nu\beta\beta$ is a known process as related to all phases in $U$ and its half life is given by 
    \begin{eqnarray}
        \left( T^{0\nu}_{1/2} \right)^{-1} =G_{0\nu}|\mathcal{M}_{0\nu}|^2 m_{ee}^2,
    \end{eqnarray}
where $G_{0\nu}$ and $\mathcal{M}_{0\nu}$ are the phase space factor and the nuclear matrix element of the $0\nu\beta\beta$, respectively. The decay rate of normal double-beta decay is simply given by $G_{0\nu}|\mathcal{M}_{0\nu}|^2$ without $m_{ee}$. The effective Majorana mass of electron neutrino mass $m_{ee}$ is the only neutrino-dependent factor in the decay width, which is given as 
	\begin{eqnarray}
	m_{ee}
	\equiv |\check{m_1}U_{e1}^2+\check{m_2}U_{e2}^2+m_3U_{e3}^2|. \label{mee1}
	\end{eqnarray}
It can be expressed more explicitly in terms of phases as
	\begin{eqnarray}
	m_{ee}^2
				&=& ~m_1^2|U_{e1}|^4+m_2^2|U_{e2}|^4+m_3^2|U_{e3}|^4  \nonumber \\
				&& +2m_1m_2\cos{2(\eta_2-\eta_1)}|U_{e1}|^2|U_{e2}|^2 \label{mee2} \\
				&& +2m_1m_3\cos{2(\eta_1+\delta)}|U_{e1}|^2|U_{e3}|^2  \nonumber \\
				&& +2m_2m_3\cos{2(\eta_2+\delta)}|U_{e2}|^2|U_{e3}|^2. \nonumber
	\end{eqnarray}

\begin{figure}[t]
\centering
\includegraphics[width=0.5\textwidth]{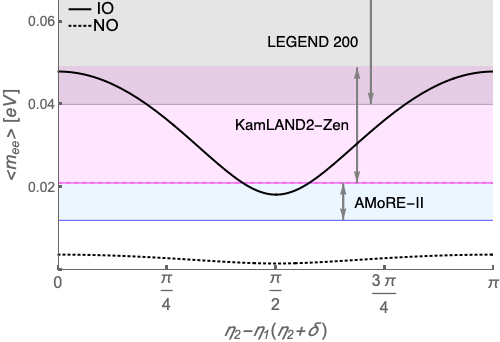}
\caption{$m_{ee}$ vs. phases. If mass ordering is hierarchical, the phase dependency is monotonous. In the effective Majorana mass, $\cos2(\eta_2+\delta)$ term is dominant for NO, while $\cos2(\eta_2-\eta_1)$ is dominant for IO. The 3$\sigma$ sensitivities of $m_{ee}$ that future experiments aim for are known as 40-73 meV for LEGEND200, 21-49 meV for KamLAND2-Zen, and 12-21 meV for AMoRE-II.}
\label{fig_mee3}
\end{figure}

The current values of neutrino masses and mixing angles and CP phase based on various oscillation experiments are as follows \cite{ParticleDataGroup:2020ssz}\cite{deSalas:2020pgw}\cite{Esteban:2020cvm}:
\begin{eqnarray}
 && \begin{array}{lclr}
        \Delta m_{21}^2 &=& (7.53\pm0.18) \times 10^{-5}~\rm{(eV)}^2 &     \\
        \sin^2\theta_{12} &=& (0.307\pm0.013) \\
        \sin^2\theta_{13} &=& (2.20\pm0.07)\times10^{-2} & \\
  \end{array} ~\begin{array}{c}\rm{NO} \\ \rm{IO}\end{array} \label{best_fit} \\ \nonumber \\
  && \begin{array}{lclc}
        \Delta m_{32}^2 &=& (2.453\pm0.033)\times10^{-3}~\rm{(eV)}^2 & \\
        \sin^2\theta_{23} &=& (0.546\pm0.021) &
  \end{array} ~\rm{NO} \label{best_fit_no} \\ \nonumber \\
  && \begin{array}{lclc}
        \Delta m_{32}^2 &=& (-2.536\pm0.034) \times 10^{-3}~\rm{(eV)}^2 &\\
        \sin^2\theta_{23} &=& (0.539\pm0.022)  &
      \end{array}  ~\rm{IO} \label{best_fit_io}\\ \nonumber \\
  && \begin{array}{lclc}
        \delta/\pi &=& 1.08+0.13 (-0.12) &  \\
        \delta/\pi &=& 1.58+0.15 (-0.16) & 
  \end{array} ~\begin{array}{ccc} & & \rm{NO} \\ & & \rm{IO}\end{array}
\end{eqnarray}
where NO means $m_1<m_2<m_3$ order and IO means $m_3<m_2<m_2$ order. Referring Eq.(\ref{mee2}), for NO, $m_{ee}$ obtains the maximum with $\eta_2+\delta=0$ or $\pi$ and the minimum with $\eta_2+\delta=\pi/2$. 
For IO, the maximum $m_{ee}$ comes up with $\eta_2-\eta_1=0$ or $\pi$ and the minimum with $\pi/2$. The curves in Fig.\ref{fig_mee3} show the aspect straightforwards.    
As upcoming $0\nu\beta\beta$ search experiments improve the sensitivity of $m_{ee}$, they can first test mass type IO and can rule out some range of $\eta_2-\eta_1$. As can be seen from the figure, the sensitivities of the known experiments do not approach to that of NO type, although NO is somehow favored by oscillation experiments \cite{T2K:2019bcf}\cite{NOvA:2019cyt}. If the neutrino mass type is NO and $0\nu\beta\beta$ is found in the experiments with sensitivity in the range 10 - 60 meV, it is necessary to modify the neutrino mass structure.  

\section{Extended seesaw mechanism}
\subsection{Simple choices for CI matrix with three right-hand neutrinos}
Canonical model of seesaw mechanism to suppress neutrino masses below 1 eV using heavy neutrino mass scale requires $SU(2)$ Higgs doublet, whose existence was experimentally confirmed \cite{ATLAS:2012yve}\cite{CMS:2012qbp}. To the following SM contents:\\
\indent - left-hand lepton doublets $L_\ell =(\nu_\ell ~\ell_L)^T_i$, \\
\indent - right-hand lepton singlets $\ell_R$,\\
\indent - Higgs doublet $H=(\phi^+ ~ \phi^0)^T$,\\
where $\ell$ is the index for $e, \mu,$ and $\tau$, non-SM singlet neutrinos $N_R$'s are added to give rise to Yukawa couplings to neutrinos, and the Majorana mass term of $N_R$ is accordingly constructed. 
After spontaneous symmetry breaking of $SU(2)$ by the vacuum expectation value $\langle \phi^0 \rangle=v$, the lepton mass terms of lagrangian reduces to
	\begin{eqnarray}
    - \mathcal{L}_\mathrm{mass} &=& v \mathcal{Y}_\ell \bar{\ell}_L \ell_R 
    + v \mathcal{Y}_\nu \bar{\nu}_\ell N^c_R \nonumber \\
    &+& \frac{1}{2} M_R \bar{N}_R N^c_R + ~\mathrm{h.c.},
	\label{lagrangian_mass}
	\end{eqnarray}
with $v= 246/\sqrt{2}~\mathrm{GeV}$.
In terms of flavors, we consider 3-generation Majorana neutrinos motivated from $SO(10)$ grand unified theory. The model is constructed on the basis where heavy Majorana neutrinos and charged leptons are represented by $N_R=(N_1 ~N_2 ~N_3)^T$ and $\ell_L=(e_L ~\mu_L ~\tau_L)^T$ and their masses $M_R$ and $\mathcal{Y}_\ell$ are diagonal. The $3\times 3$ neutrino Yukawa matrix $\mathcal{Y}_\nu$ alone is non-diagonal on the basis of $\nu_\ell = (\nu_e ~\nu_\mu ~\nu_\tau)^T$ which are the weak interaction partners of the charged leptons $\ell_L$. It is possible to express the neutrino mass term in Eq.(\ref{lagrangian_mass}) on $(\nu^T_\ell ~(N^c_R)^T)=(\nu_e ~\nu_\mu ~\nu_\tau ~N^c_1 ~N^c_2 ~N^c_3)$ such as
\begin{eqnarray}
    - \mathcal{L}_\mathrm{nu-mass} = \left(\begin{array}{cc}
    \bar{\nu}_\ell & \bar{N}_R 
    \end{array} \right)
        \left(\begin{array}{cc}
        0 & v\mathcal{Y}_\nu \\
        v\mathcal{Y}^T_\nu & M_R
        \end{array} \right)
            \left(\begin{array}{c}
                 \nu_\ell \\
                 N^c_R
            \end{array} \right),
\end{eqnarray}
where $\mathcal{Y}_\nu$ and $M_R$ are $3 \times 3$ matrices. The seesaw mechanism is a block diagonalization to separate the mass matrices of light neutrinos from heavy ones. The light Majorana masses is lifted by stepping on  heavy Majorana neutrinos such as 
    \begin{eqnarray}
        M_\nu = - v^2 \mathcal{Y}_\nu M_R^{-1} \mathcal{Y}_\nu^T,
        \label{seesaw3}
    \end{eqnarray}
which is equivalent to the mass matrix in Eq.(\ref{umu}).

Casas-Ibarra parametrization rephrases the seesaw mechanism to find the Dirac mass matrix $v\mathcal{Y}_\nu$ in terms of light Majorana massses and heavy Majorana masses such as \cite{Casas:2001sr},    
    \begin{eqnarray}
        v \mathcal{Y}_\nu = U^*\sqrt{m_L}\mathcal{O}\sqrt{M_R},
        \label{casas}
    \end{eqnarray}
where both $m_L$ and $M_R$ represent the diagonal mass matrices for light neutrinos and heavy neutrinos, respectively. The transformation matrix $U$ corresponds to the PMNS matrix. The orthogonal matrix $\mathcal{O}$ on the right-hand side is the only input that can control the structure of Yukawa matrix. In other words, a model of Yukawa matrix can be generated by the choice of an orthogonal matrix $\mathcal{O}$ in Eq.(\ref{casas}).
Three dimensional orthogonal transformation is parametrized by three rotations, $R(\varrho_{13}),~R(\varrho_{23})$, and $R(\varrho_{12})$, which were assumed to be complex in Ref.\cite{Casas:2001sr}. In this study, special choices of real angles are attempted. For example, the choice A is named for the following Dirac matrix 
    \begin{eqnarray}
        v\mathcal{Y}_A = vU^* \mathcal{O}_A  \label{yukA}
    \end{eqnarray}
with
    \begin{eqnarray}
        \mathcal{O}_A=\left(\begin{array}{ccc}
            0 & 0 & 1 \\
            0 & 1 & 0 \\
            -1 & 0 & 0 
            \end{array} \right) 
    \end{eqnarray}
obtained with $\varrho_{12}=\varrho_{23}=0$, and $\varrho_{13}=\pi/2$.
If that simple choice is made for Dirac matrix, the seesaw mechanism in Eq.(\ref{seesaw3}) reduces to the following relation
    \begin{eqnarray}
        \label{NOmass}
        \left( m_1, m_2, m_3 \right)_A
        =v^2 \left( \frac{1}{M_3}, \frac{1}{M_2}, \frac{1}{M_1} \right),
    \end{eqnarray}
and the Yukawa matrix is expressed in terms of only neutrino mixing matrix $U$ such as $\mathcal{Y}_A$=$(-U^*_{\alpha 3}$ $~e^{-i\eta_2}U^*_{\alpha 2}$ $~e^{-i\eta_1}U^*_{\alpha 1})$ with $U_{\alpha i}=(U_{ei} ~U_{\mu i} ~U_{\tau i})^T$. The above model explains that the NO mass $m_1<m_2<m_3$ is the consequence of a simple choice of $\mathcal{O}_A$ and $\mathcal{Y}_A = U^* \mathcal{O}_A$. For NO, whether the mass squared differences of Eq.(\ref{best_fit}) and Eq.(\ref{best_fit_no}) are hierarchical or quasi-degenerate, the type of difference remains open. The numerical values of light masses from Eq.(\ref{best_fit}) and Eq.(\ref{best_fit_no}) are $(m_2,~m_3)=(8.68\times10^{-3}, ~4.95\times10^{-2})~\mathrm{eV}$, and the bottom-up estimation of heavy masses is $(M_1,~M_2)=(6.06\times10^{14}, ~3.78\times10^{15})~\mathrm{GeV}$. The mass scale $M_3$ is much higher than $M_2$ so that $m_1$ can be considered close to zero. 

Another choice is constructed from the following Dirac matrix, 
    \begin{eqnarray}
        v\mathcal{Y}_B = vU^* \mathcal{O}_B \label{yukB}
    \end{eqnarray}
with
    \begin{eqnarray}
        \mathcal{O}_B=\left(\begin{array}{ccc}
            0 & 1 & 0 \\
            -1 & 0 & 0 \\
            0 & 0 & 1 
            \end{array} \right),
    \end{eqnarray}
which is an orthogonal matrix with $\varrho_{13}=\varrho_{23}=0$, and $\varrho_{12}=\pi/2$.
If Yukawa matrix is chosen as $\mathcal{Y}_B = U^* \mathcal{O}_B$ so that 
$\mathcal{Y}_B=(-e^{-i\eta_2}U^*_{\alpha 2} ~e^{-i\eta_1}U^*_{\alpha 1} ~U^*_{\alpha 3})$ with $U_{\alpha i}=(U_{ei} ~U_{\mu i} ~U_{\tau i})^T$, 
the masses of light neutrinos are determined by the heavy masses in another simple way,
    \begin{eqnarray}
        \label{IOmass}
        \left( m_1, m_2, m_3 \right)_B
        =v^2 \left( \frac{1}{M_2}, \frac{1}{M_1}, \frac{1}{M_3} \right),
    \end{eqnarray}
which describes the inverted ordering(IO) $m_3<m_1<m_2$ from the heavy Majorana mass order $M_1<M_2<M_3$.
It is worth paying attention to the mass relations in Eq.(\ref{IOmass}) that implies the quasi-degeneracy in Majorana mass scales $M_1$ and $M_2$. In IO, the $m_1$ and $m_2$ are constrained in the same order of magnitude, because $m_2$ is bounded by $\Delta m_{32}^2 = (-2.54\pm0.03) \times 10^{-3} \mathrm{eV}^2$, whereas $m_1$ is limited by $\Delta m_{21}^2 = (7.53\pm0.18) \times 10^{-5} \mathrm{eV}^2$. The quasi-degeneracy in the above IO choice appears in the relation between $M_1$ and $M_2$. The numerical values from the best-fit in Eq.(\ref{best_fit}) and Eq.(\ref{best_fit_io}) are $(m_1,~m_2)=(4.96\times10^{-2}, ~5.04\times10^{-2})~\mathrm{eV}$ and the heavy masses are $(M_1,~M_2)=(6.01\times10^{14}, ~6.10\times10^{14})~\mathrm{GeV}$. The mass scale $M_3$ is much higher than $M_2$ so that $m_3$ can be considered close to zero. The generalization of $\mathcal{O}$ with $\varrho_{ij}$s that are deviated from 0 or $\pi/2$ is also under study.

\subsection{Minimal extended seesaw mechanism}

An additional singlet neutrino $S$ is introduced such that the Lagrangian in Eq.(\ref{lagrangian_mass}) is modified to
    \begin{eqnarray}
    -\mathcal{L}_\mathrm{mass} + M_S \bar{S}^c N_R + \mathrm{h.c.},
    \end{eqnarray}
resulting in another Majorana mass $M_S$. 
Then the full contribution for neutrino masses is constructed on the basis $(\nu_\ell^T ~(N^c_R)^T ~S^c)=(\nu_e ~\nu_\mu ~\nu_\tau ~N^c_1 ~N^c_2 ~N^c_3 ~S^c)$,
    \begin{eqnarray}
    && - \mathcal{L}_\mathrm{MES} \\
    && = \left(\begin{array}{ccc}
    \bar{\nu}_\ell & \bar{N}_R & \bar{S}
    \end{array} \right)
        \left(\begin{array}{ccc}
        0 & v\mathcal{Y}_\nu & 0 \\
        v\mathcal{Y}^T_\nu & M_R & M_S^T \\
        0 & M_S &0
        \end{array} \right)
            \left(\begin{array}{c}
                 \nu_\ell \\
                 N^c_R \\
                 S^c
            \end{array} \right), \nonumber
    \end{eqnarray}
assuming that $M_S, ~M_D \ll M_R$. The scale of $1\times3$ matrix $M_S$ could be smaller or larger than that of $M_D$. It is not necessary to place restrictions on the individual elements of $M_S$ and $M_D$ for the condition that the fourth mass must be greater than the other three, as far as the low-energy phenomenological constraints are fulfilled. Taking the seesaw mechanism with respect to the heavy $M_R$ makes it possible to construct the effective mass terms on the light neutrinos $(\nu_\ell^T ~S^c)=(\nu_e ~\nu_\mu ~\nu_\tau ~S^c)$,
    \begin{eqnarray}
    M_\nu &=& 
    -\left(\begin{array}{c}
    v\mathcal{Y}_\nu \\
    M_S
    \end{array}\right) M_R^{-1}
    \left(\begin{array}{cc}
    v\mathcal{Y}_\nu^T & M_S^T
    \end{array}\right) \label{seesaw4} \\
        &=&
        \left(\begin{array}{cc}
        -v^2\mathcal{Y}_\nu M_R^{-1} \mathcal{Y}_\nu^T & 
        -v\mathcal{Y}_\nu M_R^{-1} M_S^T \\
        -v M_S M_R^{-1} \mathcal{Y}_\nu^T & 
        -M_S M_R^{-1} M_S^T
        \end{array} \right) 
    \end{eqnarray}
which represents the $4\times 4$ matrix of 3 active neutrinos and 1 sterile neutrino. In MES model, the $M_S$ may bring up an eV-order mass for $S$ field via another seesaw $ -M_S M_R^{-1} M_S^T$. It is worthwhile stressing that the determinant of $M_\nu$ vanishes, indicating that the lightest neutrino is massless \cite{Zhang:2011vh}\cite{Nath:2016mts}. The above mass matrix is diagonalized by $U'=U'_FP_3$ in Eq.(\ref{unitary4}) and Eq.(\ref{3phases}) as follows;
    \begin{eqnarray}
    m'_L &\equiv& \mathrm{Diag}\left(m_1, ~m_2, ~m_3, ~m_4 \right) \nonumber \\
        &=& U'^T M_\nu U'.
    \end{eqnarray}

Casas-Ibarra parametrization for the minimal extended seesaw mechanism can be expressed by
\begin{eqnarray}
    \left(\begin{array}{c}
         v\mathcal{Y}_\nu  \\
         M_S
    \end{array}\right) =
    \mathcal{R}U'^*\sqrt{m'_L}\mathcal{O'} \sqrt{M_R}, \label{casas4}
\end{eqnarray}
which will be called extended Dirac matrix hereafter. The $\mathcal{R}\equiv\mathrm{Diag}(1,1,1,r)$ is placed due to the different origins of $v\mathcal{Y}_\nu$ and $M_S$. 
The diagonal $\sqrt{m'_L}$ is now a $4\times4$ and $\mathcal{O'}$ is a $4\times3$ matrix that satisfies $\mathcal{O'}^T\mathcal{O'}=\mathbb{I}_{3\times3}$. On the other hand, $\mathcal{O'}\mathcal{O'}^T\neq\mathbb{I}_{4\times4}$ and so $\mathcal{O'}$ is no longer orthogonal. Similar non-orthogonal matrices were adopted in minimal seesaw model with two right-hand neutrinos for building a $3\times2$ Dirac matrix \cite{Xing:2020ald}.
A choice $A'$ is introduced by the following extended Dirac matrix:
\begin{eqnarray}
\left(\begin{array}{c}v\mathcal{Y}'_A\\  M_S 
\end{array}\right)
= v\mathcal{R} U'^*\mathcal{O}'_A \label{yukA_prime}
\end{eqnarray}
with
    \begin{eqnarray}
        \mathcal{O}'_A=\left(\begin{array}{ccc}
            0 & 0 & 0 \\
            0 & 0 & 1 \\
            0 & 1 & 0 \\
            -1 & 0 & 0 
            \end{array} \right).\label{OA_prime}
    \end{eqnarray}
    
The masses of light neutrinos are determined by substituting the extended Dirac matrix, $\mathcal{Y}'_A$=$(-U'^*_{\alpha 4}$ $~e^{-i\eta_3}U'^*_{\alpha 3}$ $~e^{-i\eta_2}U'^*_{\alpha 2})$ and $M_S/v=r(-U'^*_{s4} ~e^{-i\eta_3}U'^*_{s3} ~e^{-i\eta_2}U'^*_{s2})$  with $U_{\alpha i}=(U_{ei} ~U_{\mu i} ~U_{\tau i})^T$, into MES in Eq.(\ref{seesaw4}) 

\begin{eqnarray}
    \label{NOmass_prime}
    \left( m_1, m_2, m_3, m_4 \right)_A'=v^2 \left( 0, \frac{1}{M_3}, \frac{1}{M_2}, \frac{r^2}{M_1} \right),
\end{eqnarray}
which describes the NO-type light masses $m_2<m_3<m_4$ from the heavy Majorana masses $M_1<M_2<M_3$. Due to $\mathcal{O}'_A$, $m_1$ became the zeroed mass dictated by the vanishing determinant of $M_\nu$ in Eq.(\ref{seesaw4}). The numerical values from the best-fit are $(m_2,~m_3)=(8.68\times10^{-3}, ~4.95\times10^{-2})~\mathrm{eV}$ and the heavy neutrino scales are $(M_2,~M_3)=(6.06\times10^{14}, ~3.78\times10^{15})~\mathrm{GeV}$. The mass of the sterile neutrino $m_4$ can have an eV order with a choice of $r$ and $M_1$.

Another choice for the extended Dirac matrix in Eq.(\ref{casas4}) is
    \begin{eqnarray}
    \left(\begin{array}{c}
         v\mathcal{Y}'_B \\
         M_S 
    \end{array}\right)
    = v\mathcal{R}U'^*\mathcal{O}'_B \label{yukB_prime}
    \end{eqnarray}
with
    \begin{eqnarray}
        \mathcal{O}'_B=\left(\begin{array}{ccc}
            0 & 0 & 1 \\
            0 & 1 & 0 \\
            0 & 0 & 0 \\
            -1 & 0 & 0 
            \end{array} \right).\label{OB_prime}
    \end{eqnarray}
Then the IO-type masses of light neutrinos are determined by substituting the extended Dirac matrix, $\mathcal{Y}'_B$=($-U'^*_{\alpha 4}$ $e^{-i\eta_2}U'^*_{\alpha 2}$ $e^{-i\eta_1}U'^*_{\alpha 1}$) and $M_S/v$=$r(-U'^*_{s4}$ $e^{-i\eta_2}U'^*_{s2}$ $e^{-i\eta_1}U'^*_{s1})$ with $U_{\alpha i}$=$(U_{ei} ~U_{\mu i} ~U_{\tau i})^T$, into MES in Eq.(\ref{seesaw4}) 
    \begin{eqnarray}
        \label{IOmass_prime}
        \left( m_1, m_2, m_3, m_4 \right)_B'
        =v^2 \left( \frac{1}{M_3}, ~\frac{1}{M_2}, ~0, ~\frac{r^2}{M_1} \right),
    \end{eqnarray}
where the $m_3$ is specifically zeroed one due to $\mathcal{O}'_B$. Like the choice B in Eq.(\ref{IOmass}), the measured mass-squared differences in Eq.(\ref{best_fit_io}) require a very narrow gap between $M_2$ and $M_3$, which are $6.01\times10^{14}~\mathrm{GeV}$ and $6.10\times10^{14}~\mathrm{GeV}$, respectively.

The $r$ in Eq.(\ref{casas4}) may be either smaller or larger than one, but it has a lower bound constrained by $m_4$ and $M_1$ and an upper bound constrained by $M_2$. 
It has been shown, in Eq.(\ref{NOmass_prime}) and Eq.(\ref{IOmass_prime}), that overall order of the matrix $M_S$ does not need to be larger than the order of $M_D$ to have the $m_4$ larger than the other three masses. 

The best fits and the $3\sigma$ ranges of $|U_{e4}|^2,~|U_{\mu4}|^2, ~|U_{\tau4}|^2$ and $\Delta m_{41}^2$ are listed in the reference \cite{Gariazzo:2017fdh} as follows;
\begin{eqnarray}
\begin{array}{lcccc}
                & &\mathrm{best ~fit} & & 3\sigma ~\mathrm{CL} \\
    
    |U_{e4}|^2  & &0.0020            & & 0.0098 \sim 0.031 \\
    |U_{\mu4}|^2& &0.0015            & & 0.0060 \sim 0.026 \\ 
    |U_{\tau4}|^2& & 0.0032         &    & 0.0 \sim 0.039  \\
    \Delta m_{41}^2 & & 1.7~\mathrm{eV}^2 & & \label{prglo}
    \end{array}  
\end{eqnarray}
where the $\Delta m_{41}^2$ available in $3\sigma$ CL is not considered in this work because three separate islands of ranges exist yet in the analysis.
 The elements of $M_S$ are determined by the above experimental values using the relations in Eq.(\ref{4sines}).

\section{Leptogenesis from heavy Majorana neturino decays}

A heavy Majorana neutrino $N_i$ decays via Yukawa coupling, and its decay rate at tree level is
\begin{eqnarray}
    \Gamma_{N_i}&=&\Gamma (N_i \to \ell H) +
             \Gamma (N_i \to \bar{\ell} H^*) \nonumber \\
    &=&\frac{(\mathcal{Y}_\nu^\dagger\mathcal{Y}_\nu)_{ii}M_{i}}{8\pi},
\end{eqnarray}
where $M_i$ is the mass of $N_i$ and $\mathcal{Y}_\nu$ is the Yukawa matrix. From the interference between tree-level and one-loop level amplitudes, the CP asymmetry is estimated as
\begin{eqnarray}
\epsilon_i
 &=&\frac{\Gamma (N_i \to \ell H)
            - \Gamma (N_i \to \bar{\ell} H^*)}
        {\Gamma (N_i \to \ell H)
            + \Gamma (N_i \to \bar{\ell} H^*)} \\
 &=&\frac{\sum_{j \neq i}\mathrm{Im}[(\mathcal{Y}_\nu^\dagger\mathcal{Y}_\nu)^{2}_{ji}]}{8\pi(\mathcal{Y}_\nu^\dagger\mathcal{Y}_\nu)_{ii}}\{f(\frac{M^{2}_{j}}{M^{2}_{i}})+g(\frac{M^{2}_{j}}{M^{2}_{i}})\}
\end{eqnarray}
where $f(x)=\sqrt{x}[1-(1+x)\ln(\frac{1+x}{x})]$ and  $g(x)=\frac{\sqrt{x}}{1-x}$ indicate the contributions from the vertex correction and the self-energy correction, respectively \cite{Covi:1996wh}\cite{Roulet:1997xa}. If the difference between $M_i$ and $M_j$ is hierarchical, the self-energy part can be neglected. Though it can become dominant in case of $M_i\sim M_j$, it is out of consideration in this work. The Yukawa coupling of a Majorana neutrino itself violates the lepton number at Lagrangian level, and $\Delta L=1$ scattering of a Majorana neutrino and $\Delta L=2$ scattering, $HH \rightarrow \ell\ell$, mediated by a Majorana neutrino are also the sources of lepton number violation. 

The survival of the asymmetry from thermal washout effect is estimated from the comparison of the decay rate with the Hubble expansion rate of the Universe at temperature $T$,
    \begin{eqnarray}
    H(T)=\sqrt{\frac{4\pi^3g_*}{45}}\frac{T^2}{M_\mathrm{pl}}, \label{cosmo}
    \end{eqnarray}
where $M_\mathrm{pl}$ is the Planck scale 1.22~$\times$~$10^{19}$ GeV and $g_*$ =112 is the degree of freedom of universe with 3 Majorana neutrinos. The ratio of the decay rate to the Hubble parameter at $T=M_1$ is denoted by $K_1$, and can be estimated in terms of effective neutrino mass $\tilde{m_1}$ and equilibrium mass $m_*$ as follows,
\begin{equation}
 K_1\equiv\frac{\Gamma_{N_1}}{H(M_1)}=\frac{\tilde{m_1}}{m^*},   
\end{equation}
where
\begin{eqnarray}
    && \tilde{m_1}=\frac{(\mathcal{Y}_\nu^\dagger\mathcal{Y}_\nu)_{11}v^2}{M_{1}} \nonumber \\
    && m^*= \frac{16\pi^{5/2}}{3\sqrt{5}}\frac{\sqrt{g_*}v^2}{M_\mathrm{pl}}.
\end{eqnarray}

The observed baryon asymmetry normalized over the entropy density is given by $Y_B=(n_B-n_{\bar{B}})/s=(8.61\pm 0.05)\times 10^{-11}$ \cite{ParticleDataGroup:2020ssz}. The most well-established theory is that baryon asymmetry was produced from the lepton asymmetry through sphaleron process \cite{Harvey:1990qw}: 
\begin{equation}
    Y_{B}=\frac{a}{a-1}Y_{L}
\end{equation}
where $a=28/79$ is determined by $(8N_F+4N_H)/(22N_F+13N_H)$ for the Standard Model with the number of fermion generations $N_F=3$ and a single Higgs doublet $N_H=1$. The lepton asymmetry $Y_L$ derived from the Yukawa couplings and their interference is expressed by
    \begin{eqnarray}
    Y_{L}= \sum^3_{i=3}\kappa_{i}\frac{\epsilon_{i}}{g_*}, \label{lepto}
    \end{eqnarray}
where the dilution factor $\kappa_i$ is a function of $K$. The dilution factor, $\kappa_i$, is given by 
    \begin{eqnarray}
    \kappa_i = -\frac{0.3}{K_i(\mathrm{ln} K_i)^{3/5}},
    \end{eqnarray}
in the range $10\lesssim K_i \lesssim 10^6$ \cite{Nielsen:2001fy}\cite{Pilaftsis:1998pd}.
If Majorana neutrinos decay through the Yukawa couplings of choice A in Eq.(\ref{yukA}) or choice B in Eq.(\ref{yukB}), the CP asymmetry generated from imaginary parts in off-diagonal elements of $\mathcal{Y}^\dagger \mathcal{Y}$ cannot be produced. The unitarity of $U_\mathrm{PMNS}$ results in $\mathcal{Y}_A^\dagger \mathcal{Y}_A=\mathbb{I}$ and $\mathcal{Y}_B^\dagger \mathcal{Y}_B=\mathbb{I}$.
It is clear that $\mathrm{Im}[(\mathcal{Y}_\nu^\dagger\mathcal{Y}_\nu)^{2}_{ji}]$ vanishes as long as $\mathcal{O}$ in Eq.(\ref{casas}) does not adopt complex angles. Thus, the Davidson-Ibarra bound \cite{Davidson:2002qv} is not considered since it is built on the base that the CP-violating imaginary part in Yukawa matrix is related with complex angles in the orthogonal matrix $\mathcal{O}$.

In choice $A'$ in Eq.(\ref{yukA_prime}) and choice $B'$ in Eq.(\ref{yukB_prime}), the $3\times3$ matrices $\mathcal{Y}_A'$ and $\mathcal{Y}_B'$ avoid the unitary property of $U'_\mathrm{F}$ and non-vanishing imaginary parts of off-diagonal elements in $\mathcal{Y'}^\dagger\mathcal{Y'}$ give rise to the CP asymmetry $\epsilon_i$. When the mass relations in Eq.(\ref{NOmass_prime}) and Eq.(\ref{IOmass_prime}) are applied, the $\mathcal{Y'}^\dagger\mathcal{Y'}$ are obtained in terms of the elements of mixing matrices including both Dirac and Majorana phases as follows; 
    \begin{eqnarray}
         \mathcal{Y}_A'^\dagger \mathcal{Y}_A' &-&\mathbb{I} = \label{yty_A}\\
         &-&\left(\begin{array}{ccc}
            |U'_{s4}|^2 & U'^*_{s4}U'_{s3}e^{-i\eta_3} & U'^*_{s4}U'_{s2}e^{-i\eta_2} \\
            \checkmark & |U'_{s3}|^2 & -U'^*_{s3}U'_{s2}e^{-i(\eta_2-\eta_3)} \\
            \checkmark & \checkmark & |U'_{s2}|^2 
            \end{array} \right), \nonumber   \\
         \mathcal{Y}_B'^\dagger \mathcal{Y}_B' &-&\mathbb{I}  =  \label{yty_B} \\
         &-&\left(\begin{array}{ccc}
           |U'_{s4}|^2 & U'^*_{s4}U'_{s2}e^{-i\eta_2} & U'^*_{s4}U'_{s1}e^{-i\eta_1} \\
            \checkmark & |U'_{s2}|^2 & -U'^*_{s2}U'_{s1}e^{-i(\eta_1-\eta_2)} \\
            \checkmark & \checkmark & |U'_{s1}|^2 
            \end{array} \right).\nonumber
    \end{eqnarray}
In each choice, a phase among three Majorana phases in Eq.(\ref{3phases}) is removed along with a zero mass, and the three Dirac phases introduced in Eq.(\ref{ckm}) and Eq.(\ref{6rotation}) are imbedded in the elements of $U'_F$. The elements marked by $\checkmark$ are the complex conjugate of the transposed elements, respectively.
The simple choices predict the mass scales of Majorana neutrinos such as 
\begin{eqnarray} \label{M123}
    &&\begin{array}{rl}
    \mathrm{Choice ~A'} & \mathrm{and ~B'}\\
     & M_1= r^2(2.3\times 10^{13}) ~ \mathrm{GeV} \\
    \end{array} \\
    &&\begin{array}{rl}
    \mathrm{Choice ~A'} &  \\
     & M_2=6.1\times 10^{14} ~ \mathrm{GeV} \\
     & M_3=3.8\times 10^{15} ~ \mathrm{GeV}
    \end{array}\\
    &&\begin{array}{rl}
    \mathrm{Choice ~B'} &  \\
     & M_2=6.0\times 10^{14} ~ \mathrm{GeV} \\
     & M_3=6.1\times 10^{14} ~ \mathrm{GeV}
    \end{array} 
\end{eqnarray}
which rephrase Eq.(\ref{NOmass_prime}) and Eq.(\ref{IOmass_prime}). The $M_1$ is proportional to $r^2$ with fixed $m_4$ given in Eq.(\ref{prglo}).

\begin{figure}[t]%
\includegraphics[width=0.47\textwidth ]{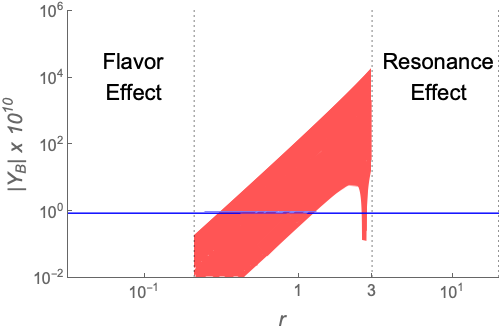} %
\includegraphics[width=0.47\textwidth ]{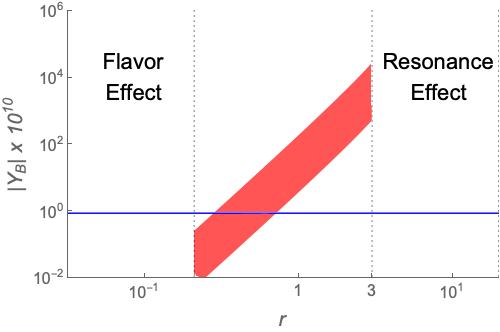} %
\caption{Baryon asymmetry's dependency on $r$ for fixed Dirac phase, $\delta=1.08\pi$ for NO(left) and $\delta=1.58\pi$ for IO(right). Other Dirac phases and Majorana phases run in full ranges. The size of $r$ has the direct correlation with $M_1$ in Eq.(\ref{M123}), and the vanilla leptogenesis model with our choices works for $0.21<r<2.95$. The blue line indicates the observed baryon asymmetry. }%
\label{fig_yB_r_NO}
\end{figure}

Thermal leptogenesis can be explained without considering the flavor effect, if the process take above $T\sim 10^{12}$ GeV \cite{Abada:2006ea}\cite{Nardi:2006fx}. The sufficiently low value of $M_1$ relative to $M_2$ and $M_3$ implies that the asymmetry generated from the decays of $N_2$ and $N_3$ should have been washed out by subsequent inverse processes, and so only the $\epsilon_1$ effectively contributed to the lepton asymmetry. In the case, $M_2\gtrsim3M_1$, only $M_1$ can contribute to $Y_{B}$, avoiding the resonant effect \cite{Xing:2020ald}. Thus, the $r$ in Eq.(\ref{M123}) should have a lower bound about 0.21 to avoid the flavor consideration and an upper bound about 5.1 not to consider resonant models. Figure~\ref{fig_yB_r_NO} shows the baryon asymmetry obtained upon the value of $r$ in our choices.

\section{light sterile neutrino as probe of leptogenesis}

The lepton asymmetry in both Choice $A'$ and Choice $B'$ is completely expressed in terms of elements of $4\times4$ unitary transformation for 3+1 light neutrinos. The survived asymmetry comes from mostly the decays of $N_1$, and its size depends on the dilution factor and the CP asymmetry as $Y_L=\kappa_1\epsilon_1/g_*$. In $K_1=K_1(\tilde{m_1})$, the $\tilde{m_1}$ is determined by $(1-|U_{s4}|^2)m_4/r^2$ using the mass relations in Eq.(\ref{NOmass_prime}) and Eq.(\ref{IOmass_prime}) and the $(\mathcal{Y^\dagger Y})_{11}$ in Eq.(\ref{yty_A}) and Eq.(\ref{yty_B}).  The CP asymmetry $\epsilon_1$ is also expressed in terms of the mixing elements of sterile neutrino as in Eq.(\ref{yty_A}) and Eq.(\ref{yty_B}). The CP asymmetry cannot exist without the sterile neutrino and its mixing angles.

The only free parameter that affects the asymmetry is $r$, the dependence of which is shown in Fig.(\ref{fig_yB_r_NO}). For example, $r=0.33$ is chosen to lead $M_1$ to the value $2.5\times 10^{12}$ GeV for the following estimation. According to the ranges in Eq.(\ref{prglo}), the $\tilde{m_1}$ is allowed in $0.08\sim 1.12$ eV, and thus the asymmetry took strong washout process because the  possible range of $K_1$ run 74 to 1028 from the 3$\sigma$ bounds. Hereafter, only best-fit values are adopted for the lepton asymmetry, because the uncertainties propagated from the $3\sigma$ ranges are exceeded by the ambiguities that arose from the additional Dirac phases $\delta_{24}$ and $\delta_{34}$ and Majorana phases $\eta_1, ~\eta_2$ and $\eta_3$. 
 
\subsection{$m_{ee}$}

\begin{figure}[t]%
\includegraphics[width=0.47\textwidth ]{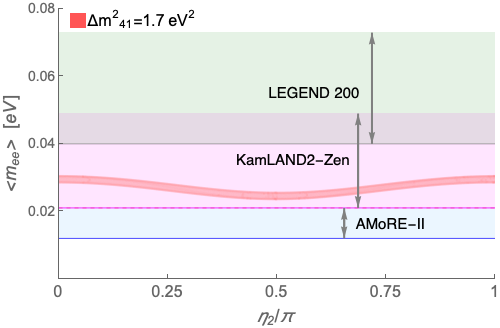} %
\includegraphics[width=0.47\textwidth ]{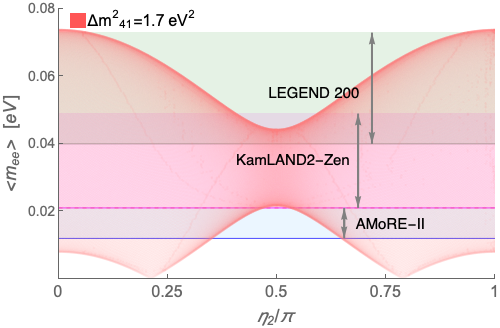}%
\caption{Effective electron-neutrino mass with 3+1 neutrinos $m_{ee}$ versus a Majorana phase $\eta_2$. (left) In NO case, running $\eta_3$ from 0 to $\pi$ results in a relatively narrow band, while $m_1=0$. (right) In IO case, running $\eta_1$ from 0 to $\pi$ results in a broad band, while $m_3=0$. }%
\label{fig_mee4}%
\end{figure}
Although most phases in MES cannot be measured by current experiments, it is possible that $m_{ee}$ measurements provide some bounds on the phases. At the end, the bounds could be tested by observed $Y_B$. For four neutrinos, $m_{ee}$ is expressed,
	\begin{eqnarray}
	|m'_{ee}|^2
				&=& ~m_1^2|U'_{e1}|^4+m_2^2|U'_{e2}|^4+m_3^2|U'_{e3}|^4+m_4^2|U'_{e4}|^4  \nonumber \\
				&& +2m_1m_2\cos{2(\eta_1-\eta_2)}|U'_{e1}|^2|U'_{e2}|^2 \nonumber \\
				&& +2m_1m_3\cos{2(\eta_1-\eta_3-\delta)}|U'_{e1}|^2|U'_{e3}|^2  \nonumber \\
				&& +2m_1m_4\cos{2\eta_1}|U'_{e1}|^2|U'_{e4}|^2  \nonumber \\
				&& +2m_2m_3\cos{2(\eta_2-\eta_3-\delta)}|U'_{e2}|^2|U'_{e3}|^2 \nonumber \\
				&& +2m_2m_4\cos{2\eta_2}|U'_{e2}|^2|U'_{e4}|^2  \nonumber \\
				&& +2m_3m_4\cos{2\eta_3}|U'_{e3}|^2|U'_{e4}|^2,  \label{mee4}
	\end{eqnarray}
which reduces to $|m_{ee}|^2$ in Eq.(\ref{mee1}) for three neutrinos. 

\textbf{Choice $A'$ for NO: } The zeroed $m_1$ removes the dependencies on $\eta_1$ in $m'_{ee}$, while it still keeps the dependencies of $\eta_2, ~\eta_3$, and $\delta$. The application of the $\delta$ in $3\sigma$ range, 0.71 to 1.99 \cite{deSalas:2020pgw}, to $m'_{ee}$ does not cause a visual effect on the band in Fig.\ref{fig_mee4}(a), where the dominant contribution come from the last two terms of Eq.(\ref{mee4}). So the figure simply shows the contribution only from both $\eta_2$ and $\eta_3$ variation 0 to $\pi$. There is a reference that showed the possibility to constrain $m_4$ and $\sin^2{2\theta_{14}}$ using the close dependency between $m_4$ and $m_{ee}$ for NO \cite{Jang:2018zug}. 
In comparison with the 3-neutrino $m_{ee}$ in Fig.\ref{fig_mee3}, the 4-neutrino $m_{ee}$ could be on the measurable range of KamLAND2-Zen and AMoRE II depending on the $\Delta m_{41}^2$ Furthermore, the 3+1 model in NO with $\Delta m_{41}^2 \geq 1.7~\mathrm{eV}^2$ is not acceptable, if AMoRE II do not find $0\nu\beta\beta$ events.

\textbf{Choice $B'$ for IO: } The zeroed $m_3$ removes the dependencies on $\eta_3$ and $\delta$ in $m'_{ee}$, while it still keeps the dependencies of $\eta_1$ and $\eta_2$. The three dominant amplitudes in Eq.(\ref{mee4}), $2m_1m_2|U'_{e1}|^2|U'_{e2}|^2, ~2m_1m_4|U'_{e1}|^2|U'_{e4}|^2$, and $2m_2m_4|U'_{e2}|^2|U'_{e4}|^2$, contribute the shape of the broad band in Fig.\ref{fig_mee4}(b). It shows the strong sensitivity on both $\eta_1$ and $\eta_2$. The three experiments mentioned in the above figures all have partial accessibility to find $0\nu\beta\beta$ events.

\subsection{$Y_B$ vs $m_{ee}$}
In four-neutrino mixing matrix, there are two additional Dirac phases $\delta_{24}$ and $\delta_{34}$, which cannot be determined by any known phenomenology either in low energy or in high energy. So the full range from 0 to $\pi$ should be covered for both $\delta_{24}$ and $\delta_{34}$. Furthermore, two Majorana phases should run almost full range until future neutrinoless double-beta decay experiments provide either positive or negative results in to reach the significant sensibility. We may seek for an indirect way in which the observation of the baryon asymmetry and the measurement of $m_{ee}$ can constrain a combination of ranges in $\delta_{24}$ and $\delta_{34}$ as well as Majorana phases.

\textbf{Choice $A'$ for NO: } Sufficient asymmetry in leptogenesis is obtained from the structure of $\mathcal{Y}_A'$ and the ratios of $M_1$ to the others. The observation of the baryon asymmetry and the discovery of neutrinoless double-beta decay can constrain the CP phases, either Dirac or Majorana. However, it is unlikely that the combined constraints significantly narrow down the ranges in $\delta_{24}$ and $\delta_{34}$, as well as those in $\eta_2$ and $\eta_3$. In Fig \ref{Yb_Mee_NO}, 9 cases are classified by combinations of $\delta_{24}$ and $\delta_{34}$, setting 0, $\pi/2$ and $\pi$ for each. As shown in the figure, a few combinations do not generate the asymmetry so as to reach the observation.
The 3$\sigma$ ranges in $|U_{\alpha 4}|^2$ in Eq.(\ref{prglo}) do not make significant changes in $Y_B$, of which effects are undercover due to sweeping of $\eta_2$ and $\eta_3$. Fig. \ref{eta2_eta3_NO} shows $Y_B$ and $m_{ee}$ in space of $\eta_2$ and $\eta_3$ for given values of $(\delta_{24}, \delta_{34})$. As $0\nu\beta\beta$ experiments improve their sensitivity, the exclusion region in $\eta_2$ and $\eta_3$ gets larger. More combinations of deltas can be rejected. For example, Fig.\ref{Yb_Mee_NO} first rules out the combinations of $(\delta_{24}, \delta_{34})$, (0,$\frac{\pi}{2}$), (0,$\pi$), ($\frac{\pi}{2}$,0), and ($\pi$,0) by $Y_B^\mathrm{obs}$, and then Fig.\ref{eta2_eta3_NO} rules out (0,0) and ($\frac{\pi}{2}$,$\pi$) by $m_{ee}=24$ meV. Only the orange area in Fig.\ref{eta2_eta3_NO} is matched to Fig.\ref{Yb_Mee_NO}.

\textbf{Choice $B'$ for IO: } Although $M_2$ and $M_3$ are almost degenerate, $M_1$ is low enough to produce the asymmetry effectively by itself and high enough to avoid the lepton flavor effect. Choice $B'$ also has the ambiguity issue from untouchable phases $\delta_{24}$ and $\delta_{34}$ as well as $\eta_1$ and $\eta_2$, as in Choice $A'$. 
Fig.\ref{Yb_Mee_IO} includes 9 cases classified by combinations of $\delta_{24}$ and $\delta_{34}$, setting 0, $\pi/2$ and $\pi$ for each, and so does Fig.~\ref{eta2_eta3_IO}. Only the orange area in Fig.\ref{eta2_eta3_IO} is matched to Fig.\ref{Yb_Mee_IO} as in Choice $A'$. Fig.\ref{Yb_Mee_IO} includes a few cases where $m_{ee}$ has a range that is not compatible with $Y_B^\mathrm{obs}$. When $(\delta_{24}, \delta_{34})$ is ($\pi$,0) or ($\pi$,$\frac{\pi}{2}$), for instance, there is no intersection between $m_{ee}=30$ meV and $Y_B^\mathrm{obs}$ in positive area of Fig. \ref{eta2_eta3_IO}. As a result, those combinations are ruled out.

\section{Concluding remarks}

We studied neutrino mass models in which three right-hand neutrinos steer the seesaw mechanism and their decays via Yukawa couplings produce the CP asymmetry from the interference between the tree level and one-loop levels. Accordingly the observed baryon asymmetry is explained such that it is converted from the lepton asymmetry by sphaleron process. The right-hand neutrinos couple with an additional light sterile neutrino besides the Yukawa couplings with active neutrinos, and so four light masses are leveraged by heavy neutrinos through the seesaw mechanism.
The model is facilitated by choosing a real orthogonal transformations in Casas-Ibarra representation of seesaw mechanism. The real transformation is given by a $4\times3$ matrix $\mathcal{O'}$ that satisfies $\mathcal{O'}^T\mathcal{O'}=\mathbb{I}_{3\times3}$ but $\mathcal{O'}\mathcal{O'}^T\neq\mathbb{I}_{4\times4}$. It has been shown that specific $\mathcal{O'_A}$ in Eq.(\ref{OA_prime}) and $\mathcal{O'_B}$ in Eq.(\ref{OB_prime}) derives certain zero masses, for instance, $m_1=0$ for NO and $m_3=0$ for IO, respectively.

It is clear that the CP asymmetry cannot be obtained if the Yukawa matrix is factorized into the unitary mixing matrix $U_\nu$ and a real orthogonal matrix $\mathcal{O}$ as in Eq.(\ref{casas}). However, if the Yukawa matrix depends on only the non-unitary partial block of a unitary $4\times4$ matrix $U_F'$ in Eq.(\ref{casas4}), then the CP asymmetry can be produced. Thus, the existence of the new Majorana field $S$ and its couplings with Majorana neutrinos pushed up the Yukawa matrix to contribute to the lepton asymmetry $Y_L$.  
The MES for four neutrinos also can be derived with only two right-hand neutrinos as in minimal seesaw models. However, the resulting light masses consist of at least two zero masses that is not acceptable in phenomenology.

Two models dictated by $\mathcal{O'}_A$ and $\mathcal{O'}_B$ construct the masses in NO and IO, respectively. The observed $Y_B^\mathrm{obs}$ and the sensitivity of $m_{ee}$ in near future have been taken to test the models. Four-neutrino global analysis provided the limits of $|U_{e4}|, ~|U_{\mu4}|, ~|U_{\tau4}|$ and $|U_{s4}|$ and there are various efforts in process to determine them. However, the improvement in precision of masses and mixing angles is eclipsed by the effect of a number of phases in four-neutrino theory. A series of figures, Fig.\ref{Yb_Mee_NO} to Fig.\ref{eta2_eta3_IO}, show that the vagueness from additional phases, either Dirac or Majorana, looks difficult to clear out. 
The feature of our choice is that high energy legacy $Y_B^\mathrm{obs}$ can be explained by low energy measurable quantities, such as masses, mixing angles and phases of four neutrinos. Still there is one free parameter $r$ we need to choose.
It is possible to find excluded ranges in $\delta_{24}$ and $\delta_{34}$ when $Y_B$ and $m_{ee}$ are inspected simultaneously.

\appendix
\section{$4\times4$ unitary transformation}  
The standard parametrization of $3\times 3$ transformation matrix for three neutrinos can be expressed such as
\begin{equation}
    V~=~R\left(\theta_{23}\right)
        R\left(\theta_{13},\delta_{13}\right)
        R\left(\theta_{12}\right) 
        =\left(\begin{array}{ccc}
        c_{12}c_{13}&s_{12}c_{13}&s_{13}e^{-i\delta}\\
        -s_{12}c_{23}-c_{12}s_{23}s_{13}e^{i\delta} &
        c_{12}c_{23}-s_{12}s_{23}s_{13}e^{i\delta} &
        s_{23}c_{13} \\
        s_{12}s_{23}-c_{12}c_{23}s_{13}e^{i\delta} &
        -c_{12}s_{23}-s_{12}c_{23}s_{13}e^{i\delta} &
        c_{23}c_{13}
    \end{array}\right)\label{ckm}
\end{equation} 
where $s_{ij}$ and $c_{ij}$ denotes $\sin{\theta_{ij}}$ and
$\cos{\theta_{ij}}$. For Majorana neutrinos, the transformation $U$ is given by the product of $V$ and $P_2$, where
\begin{eqnarray}
P_2=\left(\begin{array}{ccc}
        e^{i\eta_1} & 0 & 0 \\
         0 & e^{i\eta_2} &  0 \\
          0 & 0 & 1 
        \end{array}\right). \label{2phases}
\end{eqnarray}
The matrix $V$ is called $U_\mathrm{PMNS}$ and its elements are denoted as
\begin{eqnarray}
U_\mathrm{PMNS}=\left(\begin{array}{ccc}
        U_{e1} & U_{e2} & U_{e3} \\
         U_{\mu1} & U_{\mu2} & U_{\mu3} \\
          U_{\tau1} & U_{\tau2} & U_{\tau3}
        \end{array}\right).
\end{eqnarray}

When a sterile neutrino with $T_{3L}=0$ is added to the neutrino contents, the $4\times4$ unitary transformation can be denoted by
\begin{eqnarray}
        U_F'=\left(\begin{array}{cccc}
        U'_{e1} & U'_{e2} & U'_{e3} & U'_{e4}\\
         U'_{\mu1} & U'_{\mu2} & U'_{\mu3} & U'_{\mu4} \\
          U'_{\tau1} & U'_{\tau2} & U'_{\tau3} & U'_{\tau4}\\
           U'_{s1} & U'_{s2} & U'_{s3} & U'_{s4} 
        \end{array}\right),  \label{unitary4}
\end{eqnarray}
which consists of six rotations such as
    \begin{eqnarray} 
   U'_F &=& R\left(\theta_{34},\delta_{34}\right)
          R\left(\theta_{24},\delta_{24}\right)R\left(\theta_{14}\right)
          \left(\begin{array}{cc}
              V & 0 \\
              0 & 1
          \end{array}\right) \label{6rotation}
    \end{eqnarray}
where each $R(\theta_{ij})$ is a $4\times 4$ rotation matrix and the $V$ is defined in Eq.(\ref{ckm}). The elements of $U'_F$ are specified in terms of mixing angles and phases of the sterile neutrino: 
 \begin{eqnarray}
 &&  \left( \begin{array}{c}
        U'_{e1} \\ U'_{\mu1} \\ U'_{\tau1} \\  U'_{s1}
    \end{array} \right) 
    = \left( \begin{array}{c}
    V_{11} c_{14} \\ 
    -V_{11} s_{14}s_{24}e^{-i \delta _{24}}+V_{21} c_{24} \\ 
    -V_{11} s_{14}s_{34}c_{24}e^{-i \delta _{34}}-V_{21} s_{24}s_{34}e^{i (\delta _{24}-\delta _{34})} +V_{31} c_{34} \\  
    -V_{11} s_{14}c_{24}c_{34} - V_{21} s_{24}c_{34}e^{i \delta _{24}}-V_{31} s_{34}e^{i \delta _{34}}
    \end{array} \right) \nonumber \\
 &&  \left( \begin{array}{c}
        U'_{e2} \\ U'_{\mu2} \\ U'_{\tau2} \\  U'_{s2}
    \end{array} \right) 
    = \left( \begin{array}{c}
        V_{12} c_{14} \\ 
        - V_{12} s_{14}s_{24}e^{-i \delta _{24}}+V_{22} c_{24} \\ 
        - V_{12} s_{14}s_{34}c_{24}e^{-i \delta _{34}}- V_{22} s_{24}s_{34}e^{i (\delta _{24}-\delta _{34})}+V_{32}c_{34} \\  
        - V_{12} s_{14}c_{24}c_{34}- V_{22} s_{24} c_{34}e^{-i \delta _{24}}-V_{32} s_{34}e^{i \delta _{34}}
 \end{array} \right) \nonumber \\
 &&  \left( \begin{array}{c}
        U'_{e3} \\ U'_{\mu3} \\ U'_{\tau3} \\  U'_{s3}
    \end{array} \right) 
    = \left( \begin{array}{c}
        V_{13} c_{14} \\ 
        -V_{13}s_{14}s_{24}+V_{23} c_{24}e^{-i \delta _{24}} \\ 
        -V_{13} s_{14}s_{34}c_{24}e^{-i \delta _{34}} - V_{23}s_{24}s_{34}e^{i (\delta _{24}-\delta _{34})} +V_{33} c_{34} \\
        -V_{13} s_{14}c_{24}c_{34} -V_{23} s_{24}c_{34}e^{i \delta _{24}} -V_{33}s_{34}e^{i \delta _{34}}
 \end{array} \right) \nonumber \\
 &&  \left( \begin{array}{c}
        U'_{e4} \\ U'_{\mu4} \\ U'_{\tau4} \\  U'_{s4}
    \end{array} \right) 
    = \left( \begin{array}{c}
        s_{14} \\ 
        s_{24}c_{14}e^{-i \delta _{24}} \\ 
        s_{34}c_{14} c_{24}e^{-i \delta _{34}} \\  
        c_{14}c_{24}c_{34}
 \end{array}\right)
 \end{eqnarray}
For the transformation of Majorana neutrinos, the diagonal phase transformation with three Majorana phases also here is attached such that $U'=U'_FP_3$ with the following diagonal phase transformation:
\begin{eqnarray}
P_3=\left(\begin{array}{cccc}
        e^{i\eta_1} & 0 & 0 & 0 \\
         0 & e^{i\eta_2} & & 0 \\
          0 & 0 & e^{i\eta_3} & 0 \\
           0 & 0 & 0 & 1 \\
        \end{array}\right). \label{3phases}
\end{eqnarray}

The individual rotation angle can be expressed in terms of the elements of $U'_F$,
\begin{eqnarray}
    && s_{14}^2=|U_{e4}'|^2 \nonumber \\
    && s_{24}^2=|U_{\mu4}'|^2(1-|U_{e4}'|^2)^{-1} \label{4sines}\\
    && s_{34}^2=|U_{\tau4}'|^2(1-|U_{e4}'|^2-|U_{\mu4}'|^2)^{-1} \nonumber
\end{eqnarray}
where $|U_{s4}'|^2=1-|U_{e4}'|^2-|U_{\mu4}'|^2-|U_{\tau4}'|^2.$

\begin{figure*}[t]%
{
\includegraphics[width=0.32\textwidth ]{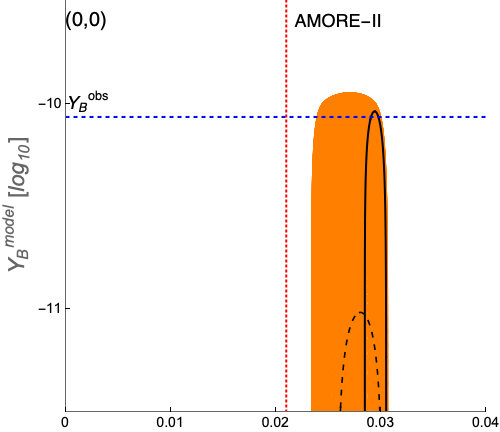} 
\includegraphics[width=0.32\textwidth ]{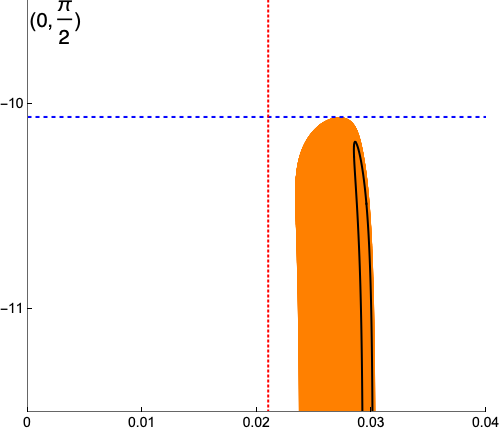}
\includegraphics[width=0.32\textwidth ]{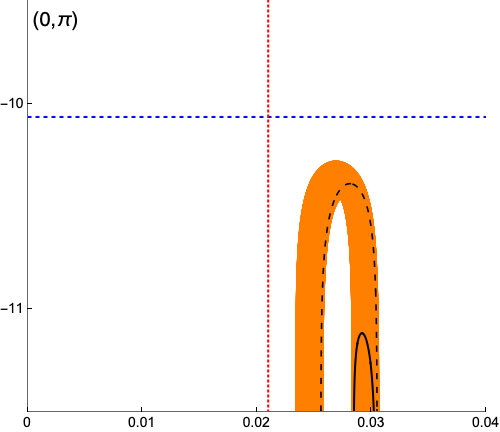}
}
{
\includegraphics[width=0.32\textwidth ]{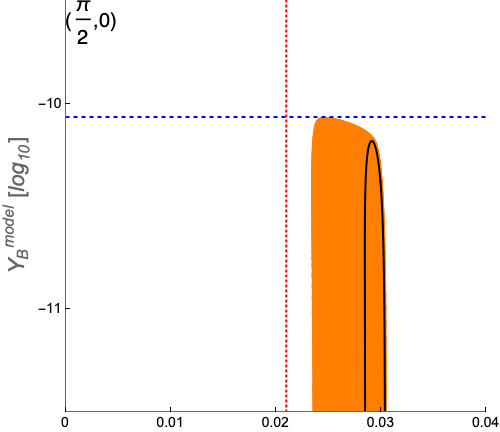} 
\includegraphics[width=0.32\textwidth ]{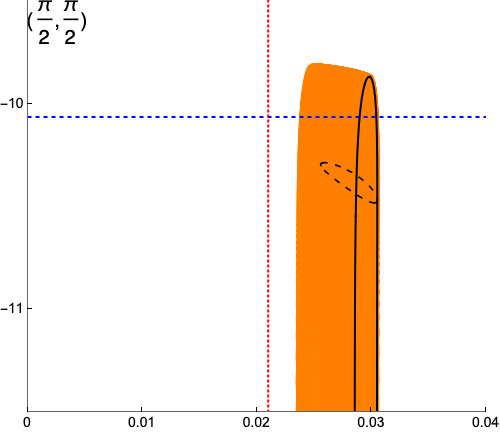}
\includegraphics[width=0.32\textwidth ]{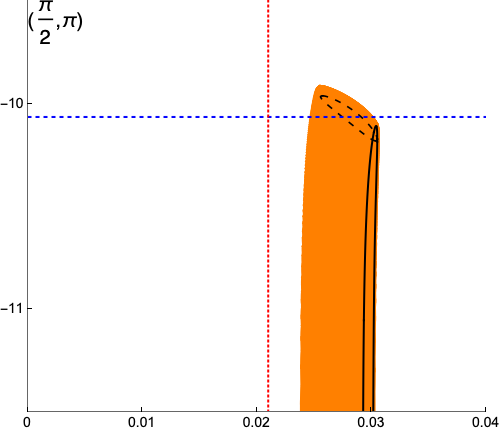}
}
{
\includegraphics[width=0.32\textwidth ]{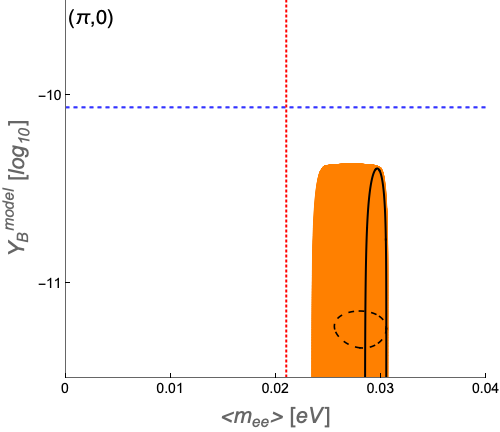} 
\includegraphics[width=0.32\textwidth ]{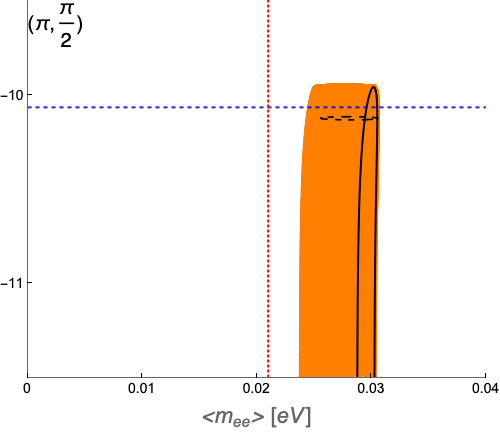}
\includegraphics[width=0.32\textwidth ]{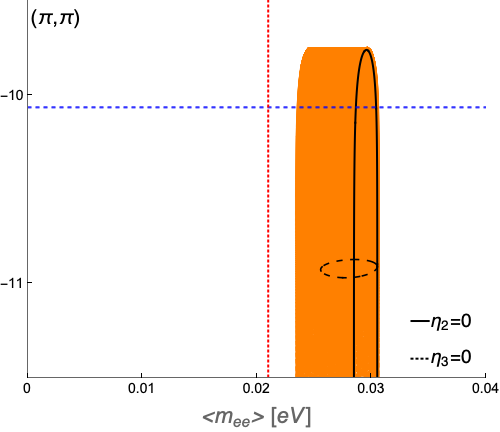}
}
\caption{$Y_B$ vs. $m_{ee}$ for given choice of $(\delta_{24} , ~\delta_{34})$ in Choice $A'$. Majorana phases $\eta_2$ and $\eta_3$ run fully 0 to $\pi$ so as to cover the colored area. The blue dotted horizontal line at $(8.61\pm 0.05)\times 10^{-11}$ indicates the observed baryon asymmetry, while the red dotted vertical line at 21 meV indicates the upper bound of sensitivity of AMoRE II. The solid and dashed lines inside each shade mark the criteria, $\eta_2=0$ and $\eta_3=0$, respectively. Only positive results are presented.}%
\label{Yb_Mee_NO} %
\end{figure*}
\begin{figure*}[t]%
{
\includegraphics[width=0.35\textwidth ]{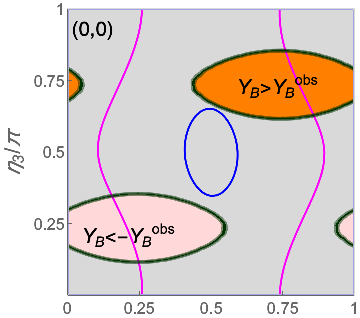} 
\includegraphics[width=0.315\textwidth ]{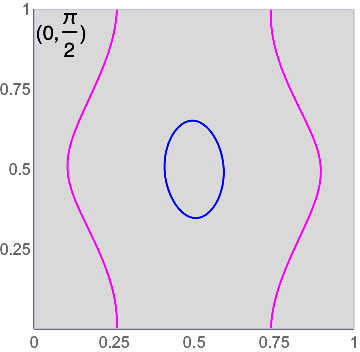}
\includegraphics[width=0.315\textwidth ]{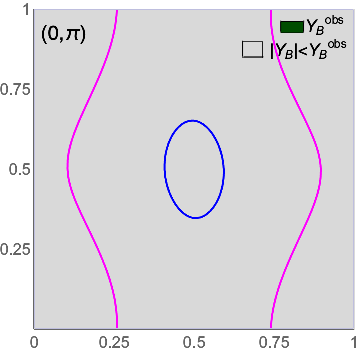}
}
{
\includegraphics[width=0.35\textwidth ]{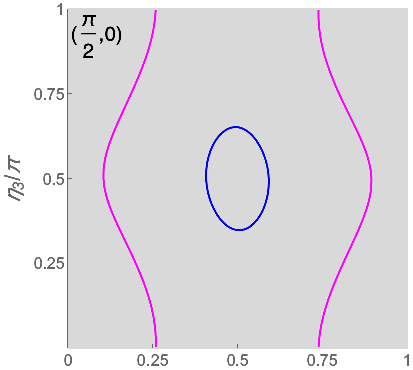} 
\includegraphics[width=0.315\textwidth ]{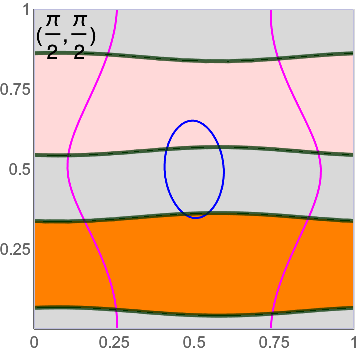}
\includegraphics[width=0.315\textwidth ]{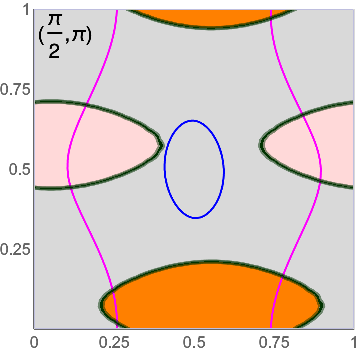}
}
{
\includegraphics[width=0.35\textwidth ]{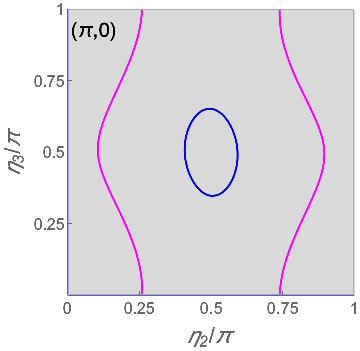} 
\includegraphics[width=0.315\textwidth ]{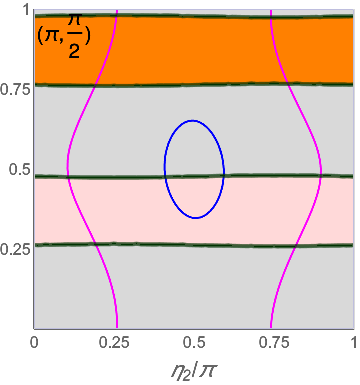}
\includegraphics[width=0.315\textwidth ]{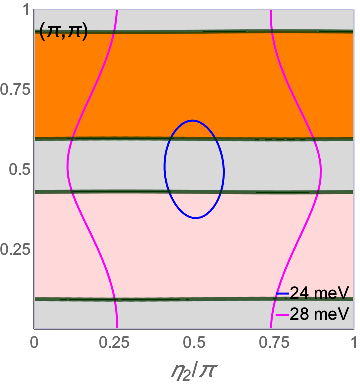}
}
\caption{Contours of $Y_B$ and $m_{ee}$ in $\eta_3-\eta_2$ space for given choice of $(\delta_{24}, ~\delta_{34})$ in Choice $A'$. The contour of $m_{ee}=28$ meV(24 meV) is represented by magenta(blue) solid line. The green curves $Y_B^\mathrm{obs}$ correspond to the contours matched to the observation. The gray areas represent $|Y_B|<Y_B^\mathrm{obs}$, while the orange and pink areas represent $Y_B>Y_B^\mathrm{obs}$ and $Y_B<-Y_B^\mathrm{obs}$, respectively.}%
\label{eta2_eta3_NO}%
\end{figure*}

\begin{figure*}[t]%
{
\includegraphics[width=0.32\textwidth ]{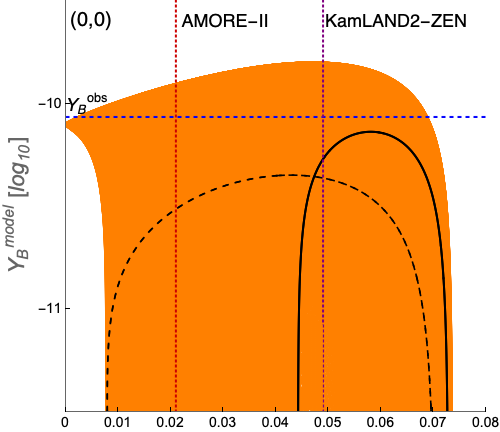} %
\includegraphics[width=0.32\textwidth ]{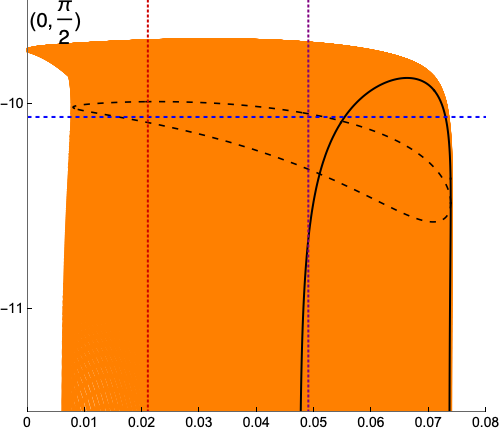}
\includegraphics[width=0.32\textwidth ]{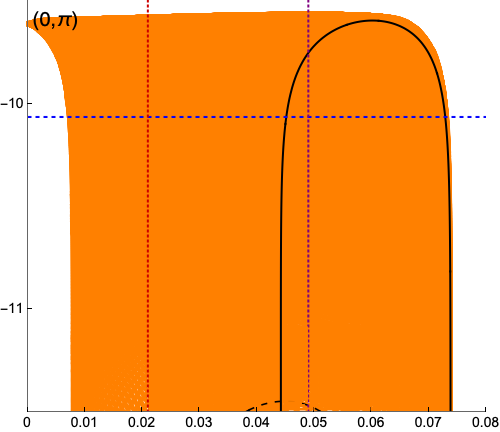}
}
{
\includegraphics[width=0.32\textwidth ]{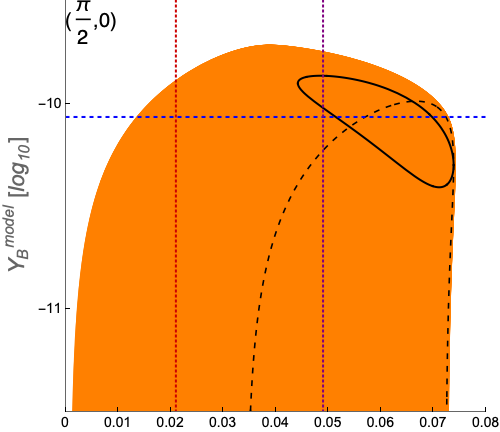} 
\includegraphics[width=0.32\textwidth ]{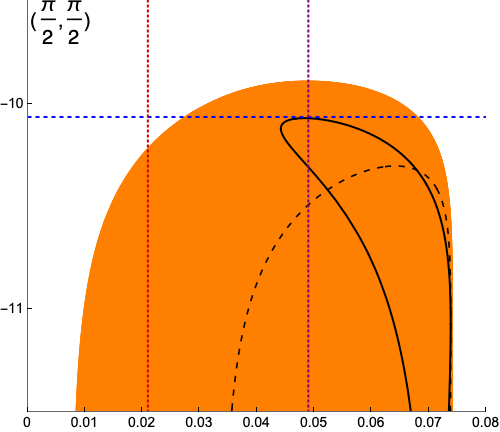} %
\includegraphics[width=0.32\textwidth ]{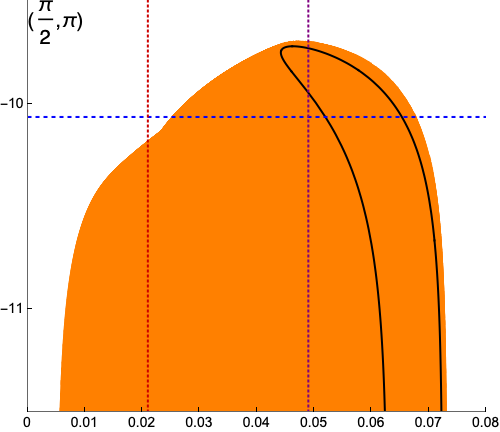}
}
{
\includegraphics[width=0.32\textwidth ]{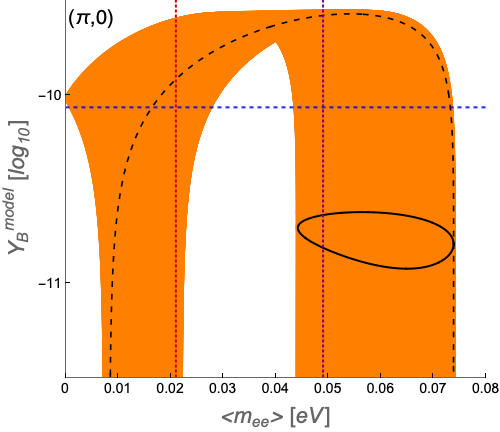} 
\includegraphics[width=0.32\textwidth ]{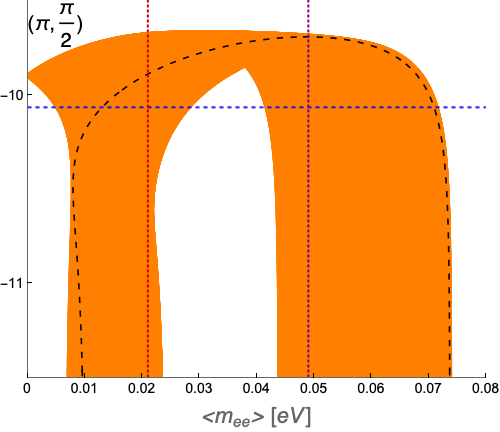} %
\includegraphics[width=0.32\textwidth ]{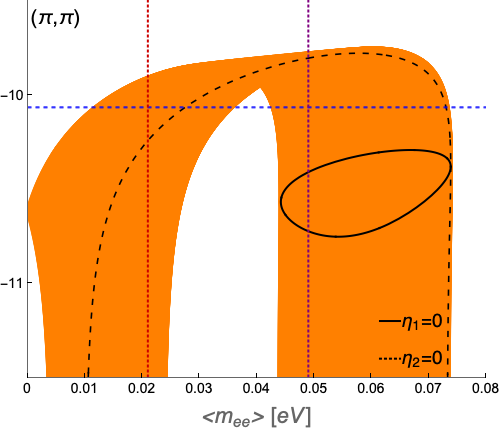}
}
\caption{$Y_B$ vs. $m_{ee}$ for given choice of $(\delta_{24} , ~\delta_{34})$ in Choice $B'$. Majorana phases $\eta_1$ and $\eta_2$ run fully 0 to $\pi$ so as to cover the colored area. The horizontal doted blue line at $(8.61\pm 0.05)\times 10^{-11}$ indicates the observed baryon asymmetry, while the vertical two lines at 21 meV and 49 meV indicate the upper bounds of sensitivities of AMoRE II and KamLAND2-Zen, respectively. The solid and dashed lines inside each shade mark the criteria $\eta_1=0$ and $\eta_2=0$, respectively. Only positive results are presented. }%
\label{Yb_Mee_IO} %
\end{figure*}

\begin{figure*}[t]%
{
\includegraphics[width=0.35\textwidth ]{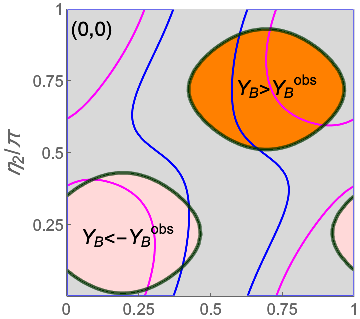} 
\includegraphics[width=0.315\textwidth ]{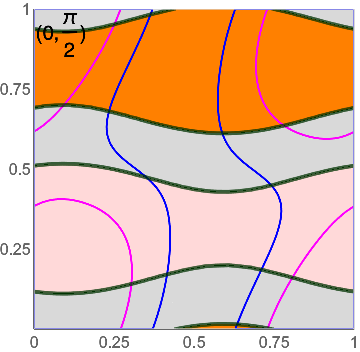}
\includegraphics[width=0.315\textwidth ]{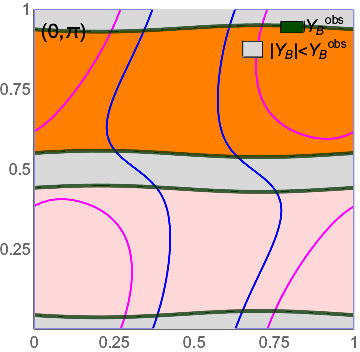}
}
{
\includegraphics[width=0.35\textwidth ]{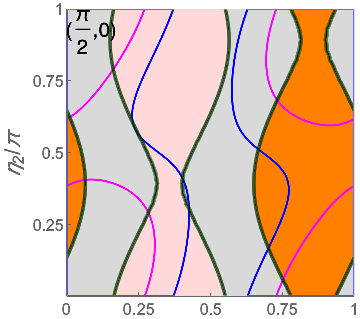} 
\includegraphics[width=0.315\textwidth ]{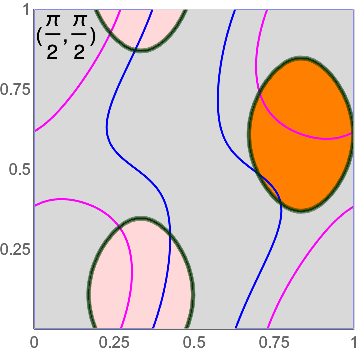}
\includegraphics[width=0.315\textwidth ]{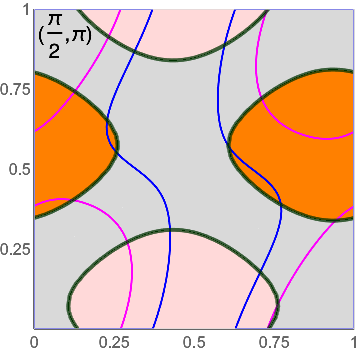}
}
{
\includegraphics[width=0.35\textwidth ]{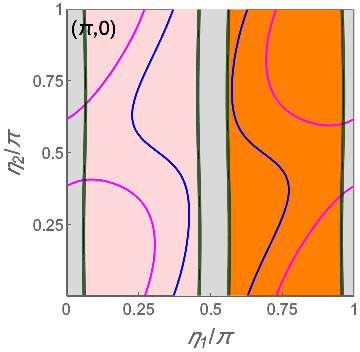} 
\includegraphics[width=0.315\textwidth ]{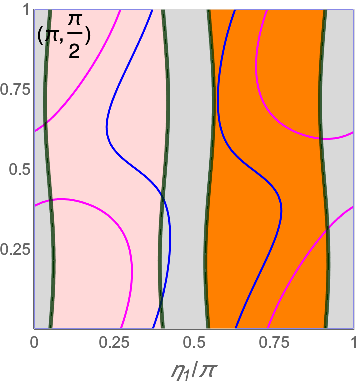}
\includegraphics[width=0.315\textwidth ]{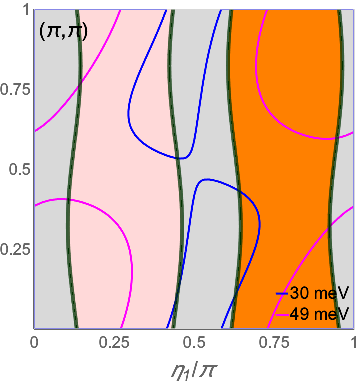}
}
\caption{Contours of $Y_B$ and $m_{ee}$ in $\eta_2-\eta_1$ space for given choice of $(\delta_{24}, ~\delta_{34})$ in Choice $B'$,  The contour of $m_{ee}=49$ meV(30 meV) is represented by magenta(blue) line. The green curves $Y_B^\mathrm{obs}$ correspond to the contours matched to the observation. The gray areas represent $|Y_B|<Y_B^\mathrm{obs}$,while the orange and pink areas represent $Y_B>Y_B^\mathrm{obs}$ and $Y_B<-Y_B^\mathrm{obs}$, respectively.}%
\label{eta2_eta3_IO}%
\end{figure*}

\acknowledgments
This work was supported by NRF grant funded by MSIT of Korea (NRF-2022R1A2C1009686) and by the Chung-Ang University research grant in 2019.

\bibliography{main}

\providecommand{\href}[2]{#2}\begingroup\raggedright\begin{thebibliography}{10}

\bibitem{DUNE:2020lwj}
{\bf DUNE} Collaboration, B.~Abi et~al., {\it {Deep Underground Neutrino
  Experiment (DUNE), Far Detector Technical Design Report, Volume I
  Introduction to DUNE}},  {\em JINST} {\bf 15} (2020), no.~08 T08008,
  [\href{http://arxiv.org/abs/2002.02967}{{\tt arXiv:2002.02967}}].

\bibitem{Hyper-Kamiokande:2018ofw}
{\bf Hyper-Kamiokande} Collaboration, K.~Abe et~al., {\it {Hyper-Kamiokande
  Design Report}},  \href{http://arxiv.org/abs/1805.04163}{{\tt
  arXiv:1805.04163}}.

\bibitem{Pontecorvo:1967fh}
B.~Pontecorvo, {\it {Neutrino Experiments and the Problem of Conservation of
  Leptonic Charge}},  {\em Zh. Eksp. Teor. Fiz.} {\bf 53} (1967) 1717--1725.

\bibitem{Maki:1962mu}
Z.~Maki, M.~Nakagawa, and S.~Sakata, {\it {Remarks on the unified model of
  elementary particles}},  {\em Prog. Theor. Phys.} {\bf 28} (1962) 870--880.

\bibitem{Agostini:2017jim}
M.~Agostini, G.~Benato, and J.~Detwiler, {\it {Discovery probability of
  next-generation neutrinoless double- \ensuremath{\beta} decay experiments}},
  {\em Phys. Rev. D} {\bf 96} (2017), no.~5 053001,
  [\href{http://arxiv.org/abs/1705.02996}{{\tt arXiv:1705.02996}}].

\bibitem{CUORE:2017tlq}
{\bf CUORE} Collaboration, C.~Alduino et~al., {\it {First Results from CUORE: A
  Search for Lepton Number Violation via $0\nu\beta\beta$ Decay of
  $^{130}$Te}},  {\em Phys. Rev. Lett.} {\bf 120} (2018), no.~13 132501,
  [\href{http://arxiv.org/abs/1710.07988}{{\tt arXiv:1710.07988}}].

\bibitem{GERDA:2018pmc}
{\bf GERDA} Collaboration, M.~Agostini et~al., {\it {Improved Limit on
  Neutrinoless Double-$\beta$ Decay of $^{76}$Ge from GERDA Phase II}},  {\em
  Phys. Rev. Lett.} {\bf 120} (2018), no.~13 132503,
  [\href{http://arxiv.org/abs/1803.11100}{{\tt arXiv:1803.11100}}].

\bibitem{Majorana:2017csj}
{\bf Majorana} Collaboration, C.~E. Aalseth et~al., {\it {Search for
  Neutrinoless Double-\ensuremath{\beta} Decay in $^{76}$Ge with the Majorana
  Demonstrator}},  {\em Phys. Rev. Lett.} {\bf 120} (2018), no.~13 132502,
  [\href{http://arxiv.org/abs/1710.11608}{{\tt arXiv:1710.11608}}].

\bibitem{EXO-200:2017hwz}
{\bf EXO-200} Collaboration, J.~B. Albert et~al., {\it {Search for nucleon
  decays with EXO-200}},  {\em Phys. Rev. D} {\bf 97} (2018), no.~7 072007,
  [\href{http://arxiv.org/abs/1710.07670}{{\tt arXiv:1710.07670}}].

\bibitem{CUPID-0:2018rcs}
{\bf CUPID-0} Collaboration, O.~Azzolini et~al., {\it {First Result on the
  Neutrinoless Double-$\beta$ Decay of $^{82}Se$ with CUPID-0}},  {\em Phys.
  Rev. Lett.} {\bf 120} (2018), no.~23 232502,
  [\href{http://arxiv.org/abs/1802.07791}{{\tt arXiv:1802.07791}}].

\bibitem{Fischer:2018squ}
{\bf SNO+} Collaboration, V.~Fischer, {\it {Search for neutrinoless double-beta
  decay with SNO+}},  in {\em {13th Conference on the Intersections of Particle
  and Nuclear Physics}}, 9, 2018.
\newblock \href{http://arxiv.org/abs/1809.05986}{{\tt arXiv:1809.05986}}.

\bibitem{NEXT:2015wlq}
{\bf NEXT} Collaboration, J.~Mart\'\i{}n-Albo et~al., {\it {Sensitivity of
  NEXT-100 to Neutrinoless Double Beta Decay}},  {\em JHEP} {\bf 05} (2016)
  159, [\href{http://arxiv.org/abs/1511.09246}{{\tt arXiv:1511.09246}}].

\bibitem{Alenkov:2019jis}
V.~Alenkov et~al., {\it {First Results from the AMoRE-Pilot neutrinoless double
  beta decay experiment}},  {\em Eur. Phys. J. C} {\bf 79} (2019), no.~9 791,
  [\href{http://arxiv.org/abs/1903.09483}{{\tt arXiv:1903.09483}}].

\bibitem{Barabash:2017sxf}
A.~S. Barabash et~al., {\it {Calorimeter development for the SuperNEMO double
  beta decay experiment}},  {\em Nucl. Instrum. Meth. A} {\bf 868} (2017)
  98--108, [\href{http://arxiv.org/abs/1707.06823}{{\tt arXiv:1707.06823}}].

\bibitem{LEGEND:2017cdu}
{\bf LEGEND} Collaboration, N.~Abgrall et~al., {\it {The Large Enriched
  Germanium Experiment for Neutrinoless Double Beta Decay (LEGEND)}},  {\em AIP
  Conf. Proc.} {\bf 1894} (2017), no.~1 020027,
  [\href{http://arxiv.org/abs/1709.01980}{{\tt arXiv:1709.01980}}].

\bibitem{KamLAND-Zen:2016pfg}
{\bf KamLAND-Zen} Collaboration, A.~Gando et~al., {\it {Search for Majorana
  Neutrinos near the Inverted Mass Hierarchy Region with KamLAND-Zen}},  {\em
  Phys. Rev. Lett.} {\bf 117} (2016), no.~8 082503,
  [\href{http://arxiv.org/abs/1605.02889}{{\tt arXiv:1605.02889}}]. [Addendum:
  Phys.Rev.Lett. 117, 109903 (2016)].

\bibitem{AMoRE:2015asn}
{\bf AMoRE} Collaboration, V.~Alenkov et~al., {\it {Technical Design Report for
  the AMoRE $0\nu\beta\beta$ Decay Search Experiment}},
  \href{http://arxiv.org/abs/1512.05957}{{\tt arXiv:1512.05957}}.

\bibitem{NEOS:2016wee}
{\bf NEOS} Collaboration, Y.~J. Ko et~al., {\it {Sterile Neutrino Search at the
  NEOS Experiment}},  {\em Phys. Rev. Lett.} {\bf 118} (2017), no.~12 121802,
  [\href{http://arxiv.org/abs/1610.05134}{{\tt arXiv:1610.05134}}].

\bibitem{PROSPECT:2018dtt}
{\bf PROSPECT} Collaboration, J.~Ashenfelter et~al., {\it {First search for
  short-baseline neutrino oscillations at HFIR with PROSPECT}},  {\em Phys.
  Rev. Lett.} {\bf 121} (2018), no.~25 251802,
  [\href{http://arxiv.org/abs/1806.02784}{{\tt arXiv:1806.02784}}].

\bibitem{DANSS:2018fnn}
{\bf DANSS} Collaboration, I.~Alekseev et~al., {\it {Search for sterile
  neutrinos at the DANSS experiment}},  {\em Phys. Lett. B} {\bf 787} (2018)
  56--63, [\href{http://arxiv.org/abs/1804.04046}{{\tt arXiv:1804.04046}}].

\bibitem{Serebrov:2020kmd}
A.~P. Serebrov et~al., {\it {Search for sterile neutrinos with the Neutrino-4
  experiment and measurement results}},  {\em Phys. Rev. D} {\bf 104} (2021),
  no.~3 032003, [\href{http://arxiv.org/abs/2005.05301}{{\tt
  arXiv:2005.05301}}].

\bibitem{SoLid:2020cen}
{\bf SoLid} Collaboration, Y.~Abreu et~al., {\it {SoLid: a short baseline
  reactor neutrino experiment}},  {\em JINST} {\bf 16} (2021), no.~02 P02025,
  [\href{http://arxiv.org/abs/2002.05914}{{\tt arXiv:2002.05914}}].

\bibitem{STEREO:2018rfh}
{\bf STEREO} Collaboration, H.~Almaz\'an et~al., {\it {Sterile Neutrino
  Constraints from the STEREO Experiment with 66 Days of Reactor-On Data}},
  {\em Phys. Rev. Lett.} {\bf 121} (2018), no.~16 161801,
  [\href{http://arxiv.org/abs/1806.02096}{{\tt arXiv:1806.02096}}].

\bibitem{RENO:2020hva}
{\bf RENO, NEOS} Collaboration, Z.~Atif et~al., {\it {Search for sterile
  neutrino oscillation using RENO and NEOS data}},
  \href{http://arxiv.org/abs/2011.00896}{{\tt arXiv:2011.00896}}.

\bibitem{LSND:2001aii}
{\bf LSND} Collaboration, A.~Aguilar-Arevalo et~al., {\it {Evidence for
  neutrino oscillations from the observation of $\bar{\nu}_e$ appearance in a
  $\bar{\nu}_\mu$ beam}},  {\em Phys. Rev. D} {\bf 64} (2001) 112007,
  [\href{http://arxiv.org/abs/hep-ex/0104049}{{\tt hep-ex/0104049}}].

\bibitem{MiniBooNE:2018esg}
{\bf MiniBooNE} Collaboration, A.~A. Aguilar-Arevalo et~al., {\it {Significant
  Excess of ElectronLike Events in the MiniBooNE Short-Baseline Neutrino
  Experiment}},  {\em Phys. Rev. Lett.} {\bf 121} (2018), no.~22 221801,
  [\href{http://arxiv.org/abs/1805.12028}{{\tt arXiv:1805.12028}}].

\bibitem{Machado:2019oxb}
P.~A. Machado, O.~Palamara, and D.~W. Schmitz, {\it {The Short-Baseline
  Neutrino Program at Fermilab}},  {\em Ann. Rev. Nucl. Part. Sci.} {\bf 69}
  (2019) 363--387, [\href{http://arxiv.org/abs/1903.04608}{{\tt
  arXiv:1903.04608}}].

\bibitem{Ajimura:2020qni}
S.~Ajimura et~al., {\it {Proposal: JSNS$^2$-II}},
  \href{http://arxiv.org/abs/2012.10807}{{\tt arXiv:2012.10807}}.

\bibitem{Kuzmin:1985mm}
V.~A. Kuzmin, V.~A. Rubakov, and M.~E. Shaposhnikov, {\it {On the Anomalous
  Electroweak Baryon Number Nonconservation in the Early Universe}},  {\em
  Phys. Lett. B} {\bf 155} (1985) 36.

\bibitem{Fukugita:1986hr}
M.~Fukugita and T.~Yanagida, {\it {Baryogenesis Without Grand Unification}},
  {\em Phys. Lett. B} {\bf 174} (1986) 45--47.

\bibitem{Luty:1992un}
M.~A. Luty, {\it {Baryogenesis via leptogenesis}},  {\em Phys. Rev. D} {\bf 45}
  (1992) 455--465.

\bibitem{Planck:2018nkj}
{\bf Planck} Collaboration, N.~Aghanim et~al., {\it {Planck 2018 results. I.
  Overview and the cosmological legacy of Planck}},  {\em Astron. Astrophys.}
  {\bf 641} (2020) A1, [\href{http://arxiv.org/abs/1807.06205}{{\tt
  arXiv:1807.06205}}].

\bibitem{Sakharov:1967dj}
A.~D. Sakharov, {\it {Violation of CP Invariance, C asymmetry, and baryon
  asymmetry of the universe}},  {\em Pisma Zh. Eksp. Teor. Fiz.} {\bf 5} (1967)
  32--35.

\bibitem{Harvey:1990qw}
J.~A. Harvey and M.~S. Turner, {\it {Cosmological baryon and lepton number in
  the presence of electroweak fermion number violation}},  {\em Phys. Rev. D}
  {\bf 42} (1990) 3344--3349.

\bibitem{Covi:1996wh}
L.~Covi, E.~Roulet, and F.~Vissani, {\it {CP violating decays in leptogenesis
  scenarios}},  {\em Phys. Lett. B} {\bf 384} (1996) 169--174,
  [\href{http://arxiv.org/abs/hep-ph/9605319}{{\tt hep-ph/9605319}}].

\bibitem{Roulet:1997xa}
E.~Roulet, L.~Covi, and F.~Vissani, {\it {On the CP asymmetries in Majorana
  neutrino decays}},  {\em Phys. Lett. B} {\bf 424} (1998) 101--105,
  [\href{http://arxiv.org/abs/hep-ph/9712468}{{\tt hep-ph/9712468}}].

\bibitem{Nielsen:2001fy}
H.~B. Nielsen and Y.~Takanishi, {\it {Baryogenesis via lepton number violation
  in anti-GUT model}},  {\em Phys. Lett. B} {\bf 507} (2001) 241--251,
  [\href{http://arxiv.org/abs/hep-ph/0101307}{{\tt hep-ph/0101307}}].

\bibitem{Xing:2020ald}
Z.-z. Xing and Z.-h. Zhao, {\it {The minimal seesaw and leptogenesis models}},
  {\em Rept. Prog. Phys.} {\bf 84} (2021), no.~6 066201,
  [\href{http://arxiv.org/abs/2008.12090}{{\tt arXiv:2008.12090}}].

\bibitem{Kang:2021stv}
S.~K. Kang, {\it {Low-energy CP violation and leptogenesis in a minimal seesaw
  model}},  {\em J. Korean Phys. Soc.} {\bf 78} (2021), no.~9 743--749.

\bibitem{Chang:2004wy}
S.~Chang, S.~K. Kang, and K.~Siyeon, {\it {Minimal seesaw model with
  tri/bi-maximal mixing and leptogenesis}},  {\em Phys. Lett. B} {\bf 597}
  (2004) 78--88, [\href{http://arxiv.org/abs/hep-ph/0404187}{{\tt
  hep-ph/0404187}}].

\bibitem{Siyeon:2016wro}
K.~Siyeon, {\it {Seesaw Scale and CP Phases in a Minimal Model of
  Leptogenesis}},  {\em J. Korean Phys. Soc.} {\bf 69} (2016), no.~11
  1638--1643, [\href{http://arxiv.org/abs/1611.04572}{{\tt arXiv:1611.04572}}].

\bibitem{Abada:2006ea}
A.~Abada, S.~Davidson, A.~Ibarra, F.~X. Josse-Michaux, M.~Losada, and
  A.~Riotto, {\it {Flavour Matters in Leptogenesis}},  {\em JHEP} {\bf 09}
  (2006) 010, [\href{http://arxiv.org/abs/hep-ph/0605281}{{\tt
  hep-ph/0605281}}].

\bibitem{Nardi:2006fx}
E.~Nardi, Y.~Nir, E.~Roulet, and J.~Racker, {\it {The Importance of flavor in
  leptogenesis}},  {\em JHEP} {\bf 01} (2006) 164,
  [\href{http://arxiv.org/abs/hep-ph/0601084}{{\tt hep-ph/0601084}}].

\bibitem{Moffat:2018smo}
K.~Moffat, S.~Pascoli, S.~T. Petcov, and J.~Turner, {\it {Leptogenesis from Low
  Energy $CP$ Violation}},  {\em JHEP} {\bf 03} (2019) 034,
  [\href{http://arxiv.org/abs/1809.08251}{{\tt arXiv:1809.08251}}].

\bibitem{Pilaftsis:2003gt}
A.~Pilaftsis and T.~E.~J. Underwood, {\it {Resonant leptogenesis}},  {\em Nucl.
  Phys. B} {\bf 692} (2004) 303--345,
  [\href{http://arxiv.org/abs/hep-ph/0309342}{{\tt hep-ph/0309342}}].

\bibitem{Casas:2001sr}
J.~A. Casas and A.~Ibarra, {\it {Oscillating neutrinos and $\mu \to e,
  \gamma$}},  {\em Nucl. Phys. B} {\bf 618} (2001) 171--204,
  [\href{http://arxiv.org/abs/hep-ph/0103065}{{\tt hep-ph/0103065}}].

\bibitem{Rodejohann:2009ve}
W.~Rodejohann, {\it {On Non-Unitary Lepton Mixing and Neutrino Mass
  Observables}},  {\em Phys. Lett. B} {\bf 684} (2010) 40--47,
  [\href{http://arxiv.org/abs/0912.3388}{{\tt arXiv:0912.3388}}].

\bibitem{Barry:2011wb}
J.~Barry, W.~Rodejohann, and H.~Zhang, {\it {Light Sterile Neutrinos: Models
  and Phenomenology}},  {\em JHEP} {\bf 07} (2011) 091,
  [\href{http://arxiv.org/abs/1105.3911}{{\tt arXiv:1105.3911}}].

\bibitem{Zhang:2011vh}
H.~Zhang, {\it {Light Sterile Neutrino in the Minimal Extended Seesaw}},  {\em
  Phys. Lett. B} {\bf 714} (2012) 262--266,
  [\href{http://arxiv.org/abs/1110.6838}{{\tt arXiv:1110.6838}}].

\bibitem{Nath:2016mts}
N.~Nath, M.~Ghosh, S.~Goswami, and S.~Gupta, {\it {Phenomenological study of
  extended seesaw model for light sterile neutrino}},  {\em JHEP} {\bf 03}
  (2017) 075, [\href{http://arxiv.org/abs/1610.09090}{{\tt arXiv:1610.09090}}].

\bibitem{Goswami:2021eqy}
S.~Goswami, V.~K.~N., A.~Mukherjee, and N.~Narendra, {\it {Leptogenesis and eV
  scale sterile neutrino}},  {\em Phys. Rev. D} {\bf 105} (2022), no.~9 095040,
  [\href{http://arxiv.org/abs/2111.14719}{{\tt arXiv:2111.14719}}].

\bibitem{Giunti:2015kza}
C.~Giunti and E.~M. Zavanin, {\it {Predictions for Neutrinoless Double-Beta
  Decay in the 3+1 Sterile Neutrino Scenario}},  {\em JHEP} {\bf 07} (2015)
  171, [\href{http://arxiv.org/abs/1505.00978}{{\tt arXiv:1505.00978}}].

\bibitem{Parke:2015goa}
S.~Parke and M.~Ross-Lonergan, {\it {Unitarity and the three flavor neutrino
  mixing matrix}},  {\em Phys. Rev. D} {\bf 93} (2016), no.~11 113009,
  [\href{http://arxiv.org/abs/1508.05095}{{\tt arXiv:1508.05095}}].

\bibitem{Gariazzo:2017fdh}
S.~Gariazzo, C.~Giunti, M.~Laveder, and Y.~F. Li, {\it {Updated Global 3+1
  Analysis of Short-BaseLine Neutrino Oscillations}},  {\em JHEP} {\bf 06}
  (2017) 135, [\href{http://arxiv.org/abs/1703.00860}{{\tt arXiv:1703.00860}}].

\bibitem{T2K:2019bcf}
{\bf T2K} Collaboration, K.~Abe et~al., {\it {Constraint on the
  matter\textendash{}antimatter symmetry-violating phase in neutrino
  oscillations}},  {\em Nature} {\bf 580} (2020), no.~7803 339--344,
  [\href{http://arxiv.org/abs/1910.03887}{{\tt arXiv:1910.03887}}]. [Erratum:
  Nature 583, E16 (2020)].

\bibitem{NOvA:2019cyt}
{\bf NOvA} Collaboration, M.~A. Acero et~al., {\it {First Measurement of
  Neutrino Oscillation Parameters using Neutrinos and Antineutrinos by NOvA}},
  {\em Phys. Rev. Lett.} {\bf 123} (2019), no.~15 151803,
  [\href{http://arxiv.org/abs/1906.04907}{{\tt arXiv:1906.04907}}].

\bibitem{ParticleDataGroup:2020ssz}
{\bf Particle Data Group} Collaboration, P.~A. Zyla et~al., {\it {Review of
  Particle Physics}},  {\em PTEP} {\bf 2020} (2020), no.~8 083C01.

\bibitem{deSalas:2020pgw}
P.~F. de~Salas, D.~V. Forero, S.~Gariazzo, P.~Mart\'\i{}nez-Mirav\'e, O.~Mena,
  C.~A. Ternes, M.~T\'ortola, and J.~W.~F. Valle, {\it {2020 global
  reassessment of the neutrino oscillation picture}},  {\em JHEP} {\bf 02}
  (2021) 071, [\href{http://arxiv.org/abs/2006.11237}{{\tt arXiv:2006.11237}}].

\bibitem{Esteban:2020cvm}
I.~Esteban, M.~C. Gonzalez-Garcia, M.~Maltoni, T.~Schwetz, and A.~Zhou, {\it
  {The fate of hints: updated global analysis of three-flavor neutrino
  oscillations}},  {\em JHEP} {\bf 09} (2020) 178,
  [\href{http://arxiv.org/abs/2007.14792}{{\tt arXiv:2007.14792}}].

\bibitem{ATLAS:2012yve}
{\bf ATLAS} Collaboration, G.~Aad et~al., {\it {Observation of a new particle
  in the search for the Standard Model Higgs boson with the ATLAS detector at
  the LHC}},  {\em Phys. Lett. B} {\bf 716} (2012) 1--29,
  [\href{http://arxiv.org/abs/1207.7214}{{\tt arXiv:1207.7214}}].

\bibitem{CMS:2012qbp}
{\bf CMS} Collaboration, S.~Chatrchyan et~al., {\it {Observation of a New Boson
  at a Mass of 125 GeV with the CMS Experiment at the LHC}},  {\em Phys. Lett.
  B} {\bf 716} (2012) 30--61, [\href{http://arxiv.org/abs/1207.7235}{{\tt
  arXiv:1207.7235}}].

\bibitem{Pilaftsis:1998pd}
A.~Pilaftsis, {\it {Heavy Majorana neutrinos and baryogenesis}},  {\em Int. J.
  Mod. Phys. A} {\bf 14} (1999) 1811--1858,
  [\href{http://arxiv.org/abs/hep-ph/9812256}{{\tt hep-ph/9812256}}].

\bibitem{Davidson:2002qv}
S.~Davidson and A.~Ibarra, {\it {A Lower bound on the right-handed neutrino
  mass from leptogenesis}},  {\em Phys. Lett. B} {\bf 535} (2002) 25--32,
  [\href{http://arxiv.org/abs/hep-ph/0202239}{{\tt hep-ph/0202239}}].

\bibitem{Jang:2018zug}
C.~H. Jang, B.~J. Kim, Y.~J. Ko, and K.~Siyeon, {\it {Neutrinoless Double Beta
  Decay and Light Sterile Neutrino}},  {\em J. Korean Phys. Soc.} {\bf 73}
  (2018), no.~11 1625--1630, [\href{http://arxiv.org/abs/1811.09957}{{\tt
  arXiv:1811.09957}}].

\end{thebibliography}\endgroup

\end{document}